# Overview of the Head and Neck Tumor Segmentation for Magnetic Resonance Guided Applications (HNTS-MRG) 2024 Challenge

Kareem A. Wahid*[1,2[0000-0002-0503-0175]], Cem Dede*[0000-0002-0543-9325], Dina M. El-Habashy[1, 3[0000-0001-5182-9976]], Serageldin Kamel[1[0000-0002-0046-4337]], Michael K. Rooney[1[0000-0002-2860-4653]], Yomna Khamis[1, 4, 5[0000-0001-8531-0849]], Moamen R. A. Abdelaal[1[0000-0003-4476-2122]], Sara Ahmed[1[0000-0002-1791-8366]], Kelsey L. Corrigan[1[0000-1111-2222-3333]], Enoch Chang[1[0000-0002-2879-5445]], Stephanie O. Dudzinski[1[0000-0001-9222-472X]], Travis C. Salzillo[1[0000-0001-6271-9879]], Brigid A. McDonald[1[0000-0003-4230-1330]], Samuel L. Mulder[1[0000-0001-5185-4805]], Lucas McCullum[1,6[0000-0001-9788-7987]], Qusai Alakayleh[1[0009-0005-0050-6567]], Carlos Sjogreen[1[0000-0002-6584-7670]], Renjie He[1[0000-0001-9166-6286]], Abdallah S.R. Mohamed[1,7[0000-0003-2064-7613]], Stephen Y. Lai[8[0000-0003-2064-7613]], John P. Christodouleas[9[0000-0001-5061-2038]], Andrew J. Schaefer[10[0000-0002-0379-741X]], Mohamed A. Naser[1**[0000-0003-1020-4966]], Clifton D. Fuller[1**[0000-0002-5264-3994]]

[1] Department of Radiation Oncology, The University of Texas MD Anderson Cancer, Houston, Texas, USA
[2] Department of Imaging Physics, The University of Texas MD Anderson Cancer, Houston, Texas, USA
[3] Transitional Year Program, Corewell Health Wiliam Beaumont, Royal Oak, MI, USA
[4] Department of Radiation Oncology, University of Maryland School of Medicine, Baltimore, MD, USA
[5] Department of Clinical Oncology and Nuclear Medicine, Faculty of Medicine, Alexandria University, Alexandria, Egypt
[6] UT MD Anderson Cancer Center UTHealth Houston Graduate School of Biomedical Sciences, Houston, USA
[7] Department of Radiation Oncology, Baylor College of Medicine, Houston, TX, USA
[8] Department of Head and Neck Surgery, The University of Texas MD Anderson Cancer, Houston, Texas, USA
[9] Elekta, Atlanta, GA
[10] Department of Computational Applied Mathematics and Operations Research, Rice University, Houston, TX, USA

* co-first authors, ** co-corresponding authors
manaser@mdanderson.org, cdfuller@mdanderson.com

**Abstract.** Magnetic resonance (MR)-guided radiation therapy (RT) is enhancing head and neck cancer (HNC) treatment through superior soft tissue contrast and longitudinal imaging capabilities. However, manual tumor segmentation remains a significant challenge, spurring interest in artificial intelligence (AI)-driven automation. To accelerate innovation in this field, we present the Head and Neck Tumor Segmentation for MR-Guided Applications (HNTS-MRG) 2024 Challenge, a satellite event of the 27th International Conference on Medical Image Computing and Computer Assisted Intervention.

This challenge addresses the scarcity of large, publicly available AI-ready adaptive RT datasets in HNC and explores the potential of incorporating multi-timepoint data to enhance RT auto-segmentation performance. Participants tackled two HNC segmentation tasks: automatic delineation of primary gross tumor volume (GTVp) and gross metastatic regional lymph nodes (GTVn) on pre-RT (Task 1) and mid-RT (Task 2) T2-weighted scans. The challenge provided 150 HNC cases for training and 50 for testing, hosted on grand-challenge.org using a Docker submission framework. In total, 19 independent teams from across the world qualified by submitting both their algorithms and corresponding papers, resulting in 18 submissions for Task 1 and 15 submissions for Task 2. Evaluation using the mean aggregated Dice Similarity Coefficient showed top-performing AI methods achieved scores of 0.825 in Task 1 and 0.733 in Task 2. These results surpassed clinician interobserver variability benchmarks, marking significant strides in automated tumor segmentation for MR-guided RT applications in HNC.

# 1    Introduction: Research Context

Radiation therapy (RT) is a cornerstone of cancer treatment for a wide variety of malignancies. Chief among the beneficiaries of RT as a treatment modality is head and neck cancer (HNC). Recent years have seen an increasing interest in MRI-guided RT planning. As opposed to more traditional computed tomography (CT)-based RT planning, MRI-guided approaches afford superior soft tissue contrast, allow for functional imaging through special multiparametric sequences (e.g., diffusion-weighted imaging), and permit daily adaptive RT through intra-therapy imaging using MRI-Linac devices [1]. Subsequently, improved treatment planning through MRI-guided adaptive RT approaches would help further maximize tumor destruction while minimizing side effects in HNC [2, 3]. Given the great potential for MRI-guided adaptive RT planning, it is anticipated that these technologies will transform clinical practice paradigms for HNC [4].

   The extensive data volume for MRI-guided HNC RT planning, particularly in adaptive settings, makes manual tumor segmentation (also referred to as contouring) by physicians — the current clinical standard — often impractical due to time constraints [5]. This is compounded by the fact that HNC tumors are among the most challenging structures for clinicians to segment [6]. Artificial intelligence (AI) approaches that leverage RT data to improve patient treatment have been an exceptional area of interest for the research community in recent years. The use of deep learning (DL) in particular has made significant strides in HNC tumor auto-segmentation [7]. These innovations have largely been driven by public data science challenges such as the HECKTOR Challenge [8] and the SegRap Challenge [9]. However, to-date, there exist no large publicly available AI-ready adaptive RT HNC datasets for public distribution. It stands to reason that community-driven AI innovations would be a remarkable asset to developing technologies for the clinical translation of MRI-guided RT.

In this public data science challenge — The Head and Neck Tumor Segmentation for MR-Guided Applications 2024 Challenge (HNTS-MRG 2024, pronounced "hunts"-"merge") — we focus on the segmentation of HNC tumors for MRI-guided adaptive RT applications. The challenge is composed of two tasks focused on automated segmentation of tumor volumes on 1) pre-RT MRI images and 2) mid-RT MRI images. An overview of HNTS-MRG 2024 is shown in **Figure 1**.

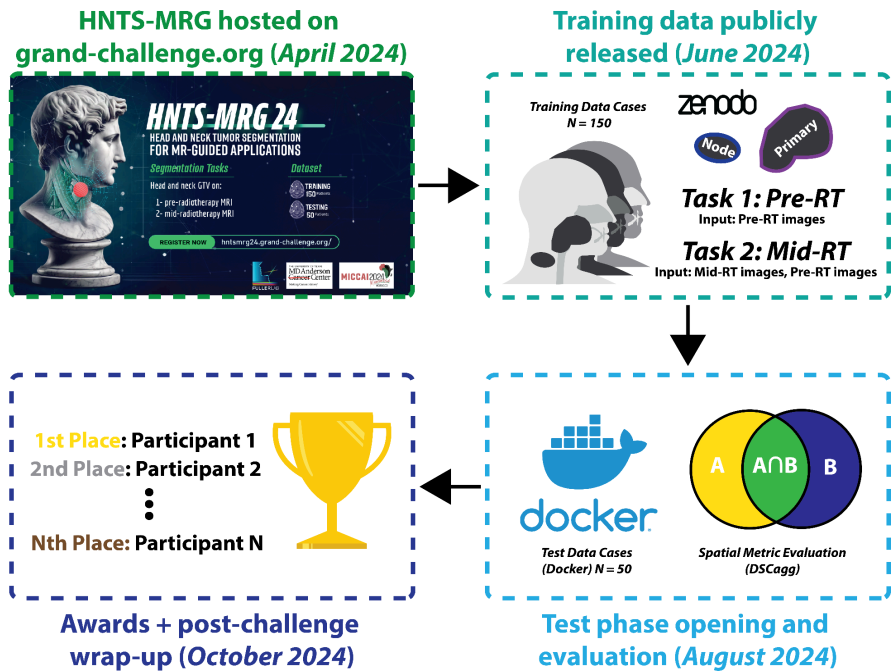

**Fig. 1.** General overview of the HNTS-MRG 2024 data science challenge. Two tasks focusing on pre-RT (Task 1) and mid-RT (Task 2) tumor segmentation using MRI scans were evaluated. Training data from 150 patients were publicly released, followed by an internal evaluation of algorithms on 50 final test patients. Subsequently, a post-challenge virtual wrap up session was held where winners were publicly announced.

## 2      Dataset and Challenge Details

### 2.1      Mission of the Challenge

**Biomedical Application.**

This data science challenge followed the Biomedical Image Analysis Challenges (BIAS) statement  reporting guidelines by Maier-Hein et al. [10] and was accepted as a satellite event for the 27th International Conference on Medical Image Computing and Computer Assisted Intervention (MICCAI). The algorithms submitted by participating teams were primarily designed for three main target applications: diagnosis, treatment planning, and medical research. These algorithms focused on two core tasks within medical imaging: segmentation and detection. Specifically, they were designed to analyze MRI images and classify/annotate individual voxels into three distinct categorical labels: primary gross tumor volume (GTVp), metastatic lymph node gross tumor volume (GTVn), or background tissue.

**Cohorts.**

Following the BIAS guidelines, we define a target cohort (i.e. subjects from whom the data would be acquired in the final biomedical application) and challenge cohort (i.e., subjects from whom challenge data were acquired). The target cohort would consist of patients with squamous cell HNC who are referred to RT planning clinics. For these patients, the automated segmentation algorithms could potentially be used directly in RT planning. The challenge cohort were patients with a confirmed histological diagnosis of squamous cell HNC who had undergone RT at our institution. This patient cohort primarily consisted of individuals with oropharyngeal cancer (OPC) or cancer of unknown primary (CUP). We included CUP patients for two key reasons. Firstly, these cases are often undetected OPC [11]. Secondly, our dataset included patients whose primary tumors had achieved complete response by mid-therapy, resulting in images where only residual mid-RT lymph nodes remained visible. This scenario closely resembles the presentation of CUP cases, making their inclusion valuable for training algorithms to detect and segment metastatic nodes across a spectrum of clinical presentations. **Section 2.3** describes further details of the challenge cohort dataset.

**Target Entity.**

*Data Origin.*

Extant  images used in this study were all acquired from the head and neck region of HNC patients. However, the exact area captured in each scan (i.e., field of view) varied somewhat between images. While some scans extended inferiorly to include parts of the lungs or superiorly to include the top of the skull, all scans consistently captured at least the area from the clavicles up to the oropharyngeal region. To make the target regions more uniform across all images, we applied a cropping technique. The specific details of this cropping method are explained in **Section 2.3**.

*Algorithm Target.*

The structures of interest for this study were GTVp and GTVn structures which are conventionally segmented by physicians for HNC RT planning. They represent the position and extent of gross tumor at the primary site and metastatic lymph nodes visible on medical imaging [12, 13].

**Task Definition.**

The challenge consisted of two tasks, pre-RT segmentation (Task 1) and mid-RT segmentation (Task 2).

Task 1 required participants to predict GTVp and GTVn tumor segmentations on unseen pre-RT scans without annotations. This is a task analogous to previous conventional tumor segmentation challenges, such as Task 1 of the 2022 HECKTOR Challenge [14] and Task 2 of the 2023 SegRap Challenge [9]. Participants were free to use mid-RT data for training their pre-RT auto-segmentation algorithms if desired.

Task 2 simulated a real-world adaptive RT scenario, providing an unseen mid-RT image alongside a pre-RT image with corresponding pre-RT segmentation. Registered and original versions of the pre-RT data would be provided during model inference (more details in **Section 2.3**). The goal was to predict GTVp and GTVn segmentations on the new mid-RT images. This task is somewhat analogous to previous challenges that utilize multiple image inputs such as the 2023 SegRap Challenge (non-contrast CT + contrast CT) [9] and the 2023 HaN-Seg Challenge (CT + MRI) [15]. To our knowledge, no previous challenges utilized patient-specific multi-timepoint MRI for segmentation purposes, making this aspect of our challenge particularly unique. Participants were free to use any combination of input images/masks to develop their mid-RT auto-segmentation algorithms.

To foster innovation while maintaining fairness, we allowed participants to leverage pre-trained model weights, foundation models, and additional external data to augment their training. However, we stipulated that all such resources must be publicly accessible and properly cited in the participants' paper submissions. This approach encouraged the use of state-of-the-art techniques while ensuring transparency and reproducibility in the challenge.

## 2.2    Online Hosting of the Challenge

HNTS-MRG 2024 was hosted on grand-challenge.org, an open-source platform that has become a de facto standard for online biomedical image analysis competitions. This platform offers essential features for running online data challenges, including an application programming interface, user management, a discussion forum, support for multi-phase competitions with separate leaderboards, and an online image results viewer, among other functionalities. Moreover, the platform utilizes Docker frameworks [16] for containerized algorithm code submissions and automated algorithm evaluation. The online webpage for HNTS-MRG 2024 was launched in April 2024, offering participants a comprehensive environment to engage in this challenge [17].

### 2.3    Challenge Cohort Dataset

**Institutional Review Board.**

Ethics approval was obtained from the University of Texas MD Anderson Cancer Center Institutional Review Board with protocol number RCR03-0800. This is a retrospective data collection protocol with a waiver of informed consent.

**Data Source.**

All data for this study were collected from a single institution: The University of Texas MD Anderson Cancer Center. T2-weighted (T2w) anatomical MRI sequences were the focus of our challenge due to their ubiquity and importance in MRI-based HNC segmentation for RT [18]. Raw T2w images in Digital Imaging and Communications in Medicine (DICOM) format were automatically extracted from a centralized institutional imaging repository (Evercore). Notably, T2w images were a mix of fat-suppressed and non-fat-suppressed images. Images include pre-RT (0-3 weeks before the start of RT) and mid-RT (2-4 weeks intra-RT) scans. No exogenous contrast enhancement agents were used for these scans. All patients were immobilized using a thermoplastic mask to aid in consistent anatomical positioning. Pre-RT and mid-RT image pairs for a given patient were consistently either fat-suppressed or non-fat-suppressed. In total, data from 202 squamous cell HNC patients were curated. T2w images of the head and neck region were acquired using a range of imaging devices and protocols. Images were acquired on the following devices: 1.5T Siemens Aera (n = 297), 1.5T Elekta Unity (n = 78), 3T Siemens Magnetom Vida (n = 19), 1.5T Siemens Magnetom Sola Fit (n = 10). A full list of imaging protocols are described in **Table 1**. Examples of T2w images for two patients are shown in **Figure 2**.

**Table 1.** Magnetic resonance imaging acquisition parameters for this study. Median values with ranges shown. Values are calculated across the entire datasets for all timepoints (pre- and mid-radiotherapy).

| Parameter | Median (range) |
|---|---|
| Repetition Time (ms) | 4800 (1400-6250) |
| Echo Time (ms) | 80 (74-375) |
| In-plane Resolution (mm) | 0.5 (0.4-0.98) |
| Slice Thickness (mm) | 2.0 (1.0-2.5) |
| Slice Gap (mm) | 2.0 (1.0-2.5) |
| Number Of Axial Slices | 120 (80-300) |
| Field Of View (mm) | 256x256 (256-520) |
| Number of Averages | 1 (1-2) |

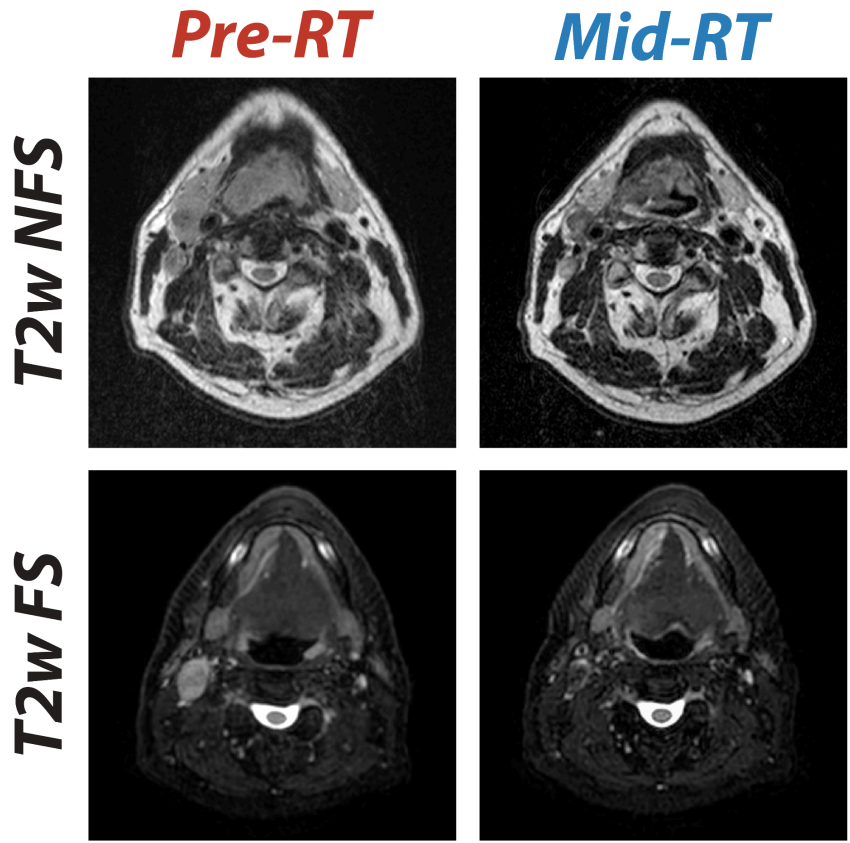

**Fig. 2.** Comparison of T2-weighted (T2w) MRI scans before radiotherapy (pre-RT) and during radiotherapy (mid-RT), showing images without fat suppression (T2w non fat suppressed [NFS], top row) and with fat suppression (T2w fat suppressed [FS], bottom row). Pre-RT scans are co-registered to the corresponding mid-RT scans.

**Annotation Characteristics.**

Each MRI scan was annotated for GTVp (maximum one per patient, potentially zero) and GTVn (variable number per patient, potentially zero). A team of 3 to 4 expert physicians each independently segmented these structures for all pre-RT and mid-RT cases. This approach aligns with recent findings from our group, suggesting that a minimum of 3 annotators is necessary to produce acceptable segmentations in these structures [19, 20] when combined using the simultaneous truth and performance level estimation (STAPLE) algorithm [21].

13 unique annotators independently contributed segmentation annotations to this study. All annotators were medical doctors with at least two years of experience in head and neck cancer segmentation. All annotators had access to patient medical histories and any previous relevant imaging (e.g., diagnostic positron emission tomography (PET)/CT imaging) via the patient's chart. Annotators were instructed to segment targets as they would normally in their clinical workflows. For mid-RT segmentations, annotators were permitted to use their registered pre-RT segmentations as a reference if desired. Segmentations were generated in Velocity AI (v.3.0.1; Varian Medical Systems; Palo Alto, CA, USA) and Raystation (v.11; RaySearch Laboratories, Stockholm, Sweden) using American Association of Physicists in Medicine Task Group 263 nomenclature [22]. A senior radiation oncology faculty member with over 15 years of experience (C.D.F.) performed final quality verification of the segmentations, where annotators were instructed to modify certain segmentations if needed (e.g., in the case of missing a lymph node).

The STAPLE algorithm implementation in SimpleITK [23] was used to combine individual segmentations into a final ground truth (also referred to as reference standard) segmentation for each case (**Figure 3**). In exceptional cases where significant discrepancies arose among annotators—such as disagreements over multiple nodal volumes or conflicting assessments of complete versus non-complete response—we deferred to the expert judgment of the senior faculty member (C.D.F.) for generating the final segmentation. The resulting ground truth label mask uses three values: 0 for background, 1 for GTVp, and 2 for GTVn (with multiple lymph nodes consolidated into a single label). An example of an image with the aforementioned labeling scheme is shown in **Figure 4**.

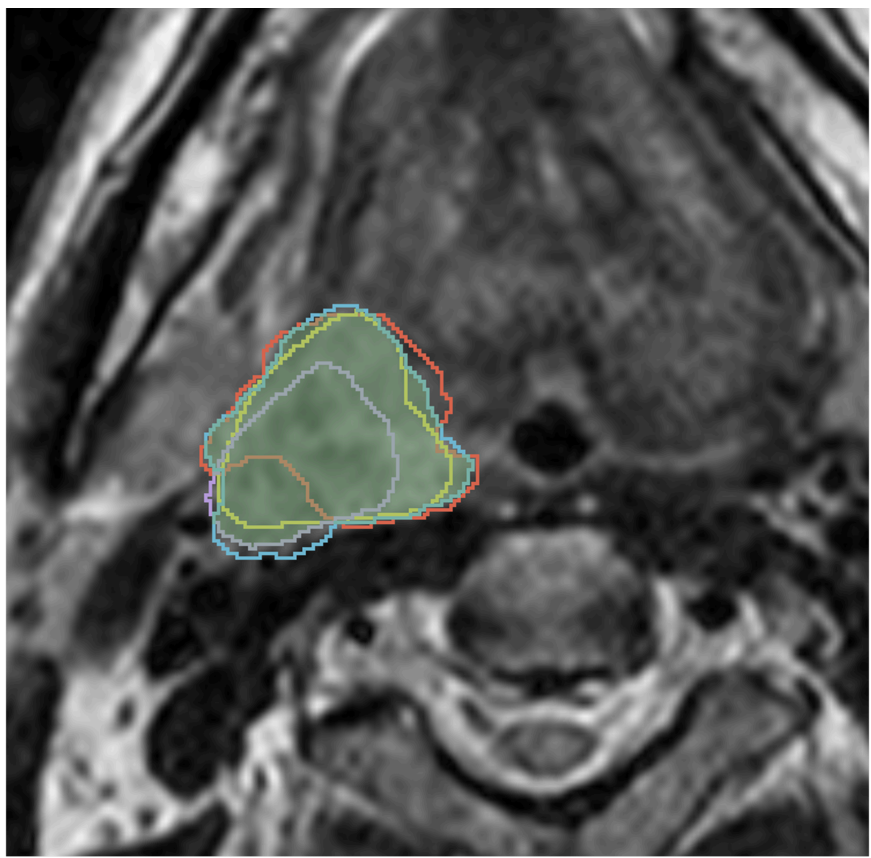

**Observer 1**, **Observer 2**, **Observer 3**, **Observer 4**,
**STAPLE Consensus**

**Fig. 3.** Example of the Simultaneous Truth and Performance Level Estimation (STAPLE) algorithm consensus process combining multiple independent annotator segmentations (red, yellow, blue, purple outlined structures) into a single final consensus segmentation (green filled in structure) for a primary gross tumor volume.

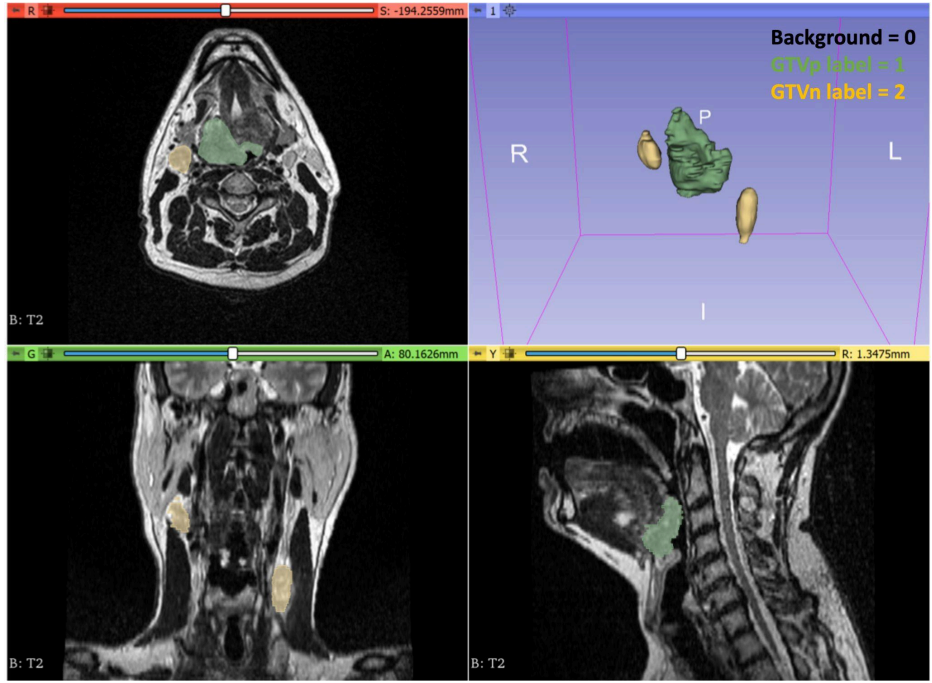

**Fig. 4.** A visual example of the mask labeling scheme for this challenge. Background = 0, primary gross tumor volume (GTVp) = 1 (green overlay), metastatic lymph node (GTVn) = 2 (yellow overlay). Masks shown are consensus segmentations from multiple independent annotators. Visualization performed in 3D Slicer.

**Data Preprocessing Methods.**

Anonymized DICOM files (MRI image and structure files) were converted to Neuroimaging Informatics Technology Initiative (NIfTI) format for ease of use by participants. Conversions were performed using DICOMRTTool v. 1.0 [24]. We chose the NIfTI format for our data due to its widespread adoption, standardized structure, and its compatibility with a broad range of analysis tools commonly used in medical imaging challenges [25]. All images were cropped from the top of the clavicles to the bottom of the nasal septum (oropharynx region to shoulders) by using manually selected inferior/superior axial slices. This allowed for more consistent image fields of view and removal of identifiable facial structures (i.e., eyes, nose, ears); cropping did not impact any of the segmented volumes.

Registered data (i.e., for Task 2) were generated using SimpleITK [23], where the mid-RT image served as the fixed image and the pre-RT image served as the moving image. Specifically, we utilized the following steps: 1. Apply a centered transformation, 2. Apply a rigid transformation, 3. Apply a deformable transformation with Elastix using a preset parameter map (Parameter map 23 in the Elastix Model Zoo which utilizes a B-spline transformation [26]). This particular deformable transformation was selected as it is open-source and was benchmarked in a previous similar application [27]. In a small minority of cases where excessive warping occurred during deformable registration, we defaulted to using only the rigid transformation. To ensure transparency and reproducibility, we provided a detailed example of our registration process on our GitHub repository [28].

**Sources of Errors - Interobserver Variability.**

The largest source of error naturally emerges from differences in annotator segmentations. We mitigated this by combining segmentations via the STAPLE algorithm which we have shown can yield acceptable segmentations given a minimal number of annotator inputs [19].

　　We evaluated the interobserver variability (IOV) for GTVp and GTVn structures in our dataset using traditional geometric measures. For each structure (i.e., GTVp and GTVn) and each timepoint (i.e., pre-RT and mid-RT), we calculated pairwise IOV. To calculate pairwise IOV, for each patient, metrics for all possible pairwise combinations between available annotator segmentations were calculated followed by computing the median value across all combinations. Naturally, the patient cases where only the senior faculty member observer contributed segmentations due to significant discrepancies among observers (see **Annotation Characteristics**) were excluded. Our analysis included four metrics: Dice Similarity coefficient (DSC), 95% Hausdorff distance (HD95), average surface distance (ASD), and surface DSC at a 2mm tolerance (SDSC). Metrics were calculated using the Surface Distances Python package [29] and in-house Python code. Pre-RT IOV DSC values showed median (interquartile range) of 0.747 (0.165) for GTVp and 0.845 (0.070) for GTVn. Mid-RT IOV DSC values were 0.558 (0.272) for GTVp and 0.808 (0.118) for GTVn. IOV based on all geometric measures are shown in **Figure 5**.

　　To provide a more comprehensive assessment of IOV relevant to our challenge, we extended our analysis beyond traditional geometric measures to include aggregated Dice Similarity Coefficient [30] (DSCagg, see **Section 2.4**) IOV calculations. We first computed intermediate metrics in a pairwise fashion across all observers for each patient case independently. These metrics were then aggregated across all cases for each unique annotator pair, enabling the calculation of DSCagg-GTVp, DSCagg-GTVn, and subsequently DSCagg-mean (the mean of DSCagg-GTVp and DSCagg-GTVn) for each annotator pair. Finally, we derived the overall IOV DSCagg values by calculating a weighted average of these annotator pair DSCagg values. The weighting factor was based on the number of cases compared for each annotator pair, ensuring appropriate annotator representation in the final score. As before, cases where only the senior faculty member observer contributed final segmentations were excluded. Final weighted IOV DSCagg values for pre-RT segmentations were DSCagg-mean = 0.806, DSCagg-GTVp = 0.757, DSCagg-GTVn = 0.854. Final weighted IOV DSCagg values for mid-RT segmentations were DSCagg-mean = 0.714, DSCagg-GTVp = 0.600, DSCagg-GTVn = 0.828.

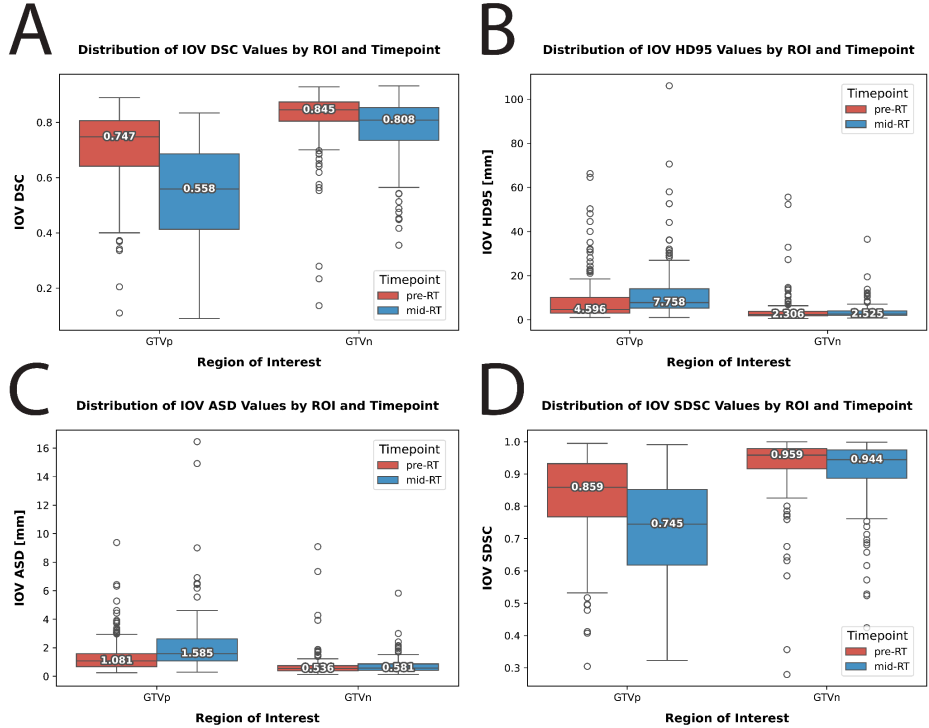

**Fig. 5.** Interobserver variability (IOV) data for primary gross tumor volume (GTVp) and nodal gross tumor volume (GTVn) regions of interest stratified by pre-radiotherapy (pre-RT) and mid-radiotherapy (mid-RT) timepoints. (A) Dice Similarity coefficient (DSC), (B) 95% Hausdorff distance (HD95), (C) average surface distance (ASD), (D) surface DSC at a 2mm tolerance (SDSC). Each datapoint corresponds to the median metric value across all pairs of observers for a given patient image. Each box represents the interquartile range, with the horizontal line indicating the median score. Outliers are shown as individual points outside the whiskers. Higher values indicate greater agreement for DSC and SDSC, while lower values indicate greater agreement for HD95 and ASD.

**Training and Test Case Characteristics.**

Tasks 1 and 2 share a common training dataset, consisting of 150 patient cases, ensuring consistency across both challenges. Training data were publicly released on Zenodo [31] under a CC BY 4.0 license. For each patient case, we provided a comprehensive set of data in NIfTI format. This dataset included six files per patient: the original pre-treatment T2-weighted MRI volume with its corresponding segmentation mask, the original mid-treatment T2-weighted MRI volume with its corresponding segmentation mask, and a registered version of the pre-treatment T2-weighted MRI volume with its corresponding segmentation mask (more details on registration in **Data Preprocessing Methods**). Each of these six files was linked to a unique anonymized case identifier, ensuring that all data for a given patient could be easily accessed and correctly associated.

The held-out private evaluation data comprised 52 additional cases, with two cases used for the challenge's preliminary debugging phase, leaving 50 cases for the final test phase (more information in **Section 2.5**). Only the challenge organizers had access to the ground truth segmentations for the test cases until final publication of the full dataset.

Training and held-out private evaluation data were partitioned to contain similar distributions based on dataset characteristics such as image fat-suppression status, tumor response, and staging. The distributions based on various parameters are shown in **Figure 6**.

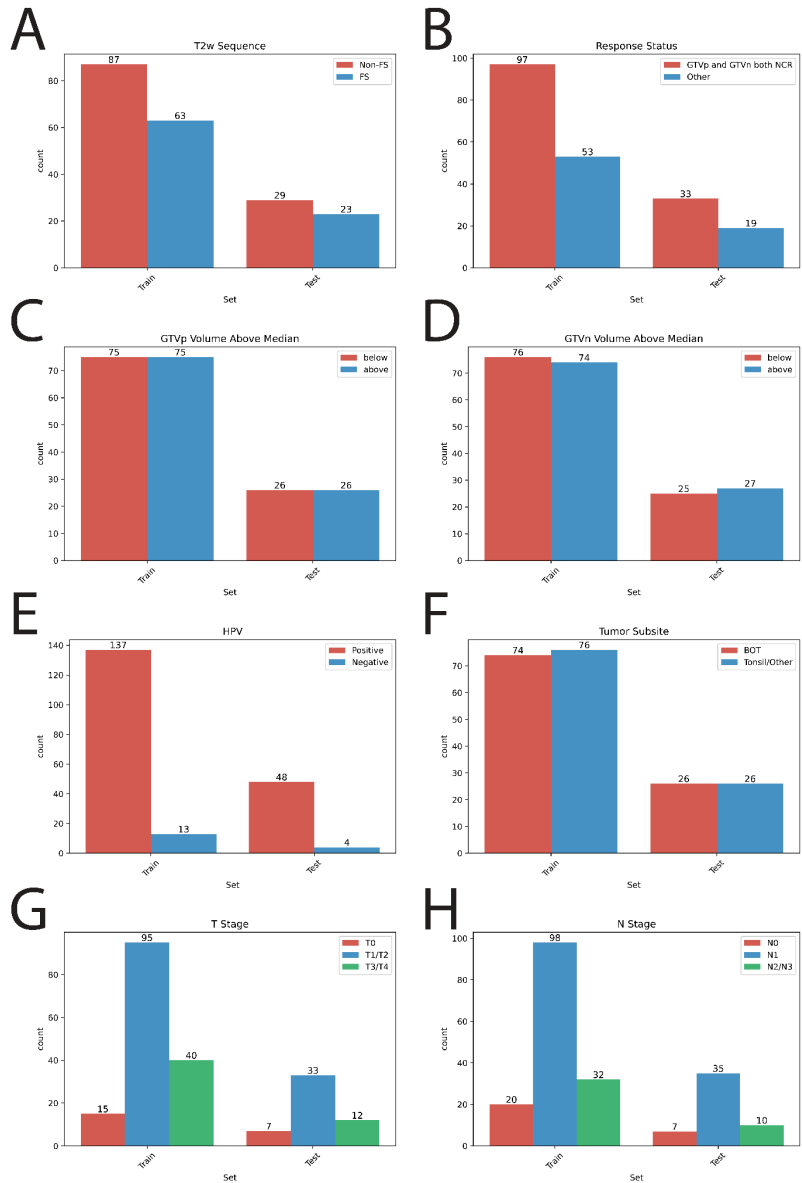

**Fig. 6.** Distribution of key parameters in training and held-out private evaluation sets (written as test). (A) T2-weighted MRI sequence type (Non-FS: non-fat suppressed, FS: fat suppressed). (B) Tumor response status at mid-therapy (NCR: non-complete response, Other: any combination of complete and non-response of primary and node). (C) Primary gross tumor volume (GTVp) and (D) nodal gross tumor volume (GTVn), both categorized as above or below the dataset median. (E) Human papillomavirus (HPV) status. (F) Tumor anatomic subsite (BOT: base of tongue). (G) T-stage and (H) N-stage as per the eighth edition American Joint Committee on Cancer staging system.

### 2.4    Assessment Method

Both tasks were evaluated in the same general manner using the aggregated Dice Similarity Coefficient (DSCagg). DSCagg was employed by Andrearczyk et al. for the segmentation task of the 2022 edition of the HECKTOR Challenge [14]. Specifically, the DSCagg metric is defined as:

$$DSCagg = \frac{2\sum_i |Ai \cap Bi|}{\sum_i |Ai| + |Bi|} \tag{1}$$

where $Ai$ and $Bi$ are the ground truth and predicted segmentation for image $i$, where $i$ spans the entire test set. Namely, DSCagg calculates intermediate metrics for each case individually and then aggregates the measurements across the test set (i.e., yields one value). DSCagg was initially described in detail by Andrearczyk et al. [30].

Conceptually, the 2022 edition of the HECKTOR Challenge had similar segmentation outputs (i.e., GTVp and GTVn for HNC patients) as our proposed challenge, so we deem DSCagg an appropriate metric. Since the presence of GTVp and GTVn were not consistent across all cases, the proposed DSCagg metric is well-suited for this task. Unlike conventional volumetric DSC, which can be overly sensitive to false positives when the ground truth mask is empty — resulting in a DSC of 0 — the DSCagg metric is more robust. It effectively handles cases where certain structures may or may not be present, providing a more balanced evaluation across diverse scenarios encountered in our data. Notably, DSCagg was shown to be a stable metric with respect to final ranking from a secondary analysis of the HECKTOR 2021 results [32], further highlighting its appropriateness for the challenge.

The metric was computed individually for GTVp (DSCagg-GTVp) and GTVn (DSCagg-GTVn), and the mean average of the two (DSCagg-mean) was used for the final challenge ranking (similar to HECKTOR 2022). The metric was calculated for Task 1 (pre-RT segmentation) and Task 2 (mid-RT segmentation) separately. We provided an example of how the DSCagg was calculated for this challenge on our GitHub repository [28].

### 2.5    Docker Submission and Challenge Phases

Algorithm submissions for the challenge were managed through grand-challenge.org, with participants required to submit their solutions as Docker container images [16]. For algorithm submissions, Task 1 participants received only an unseen pre-RT image, while Task 2 participants were provided with an unseen mid-RT image, a pre-RT image with its corresponding segmentation, and a registered pre-RT image with its corresponding registered segmentation. Toy examples of algorithm inputs for both tasks are shown in **Figure 7**. By utilizing a Docker framework, we maintained data integrity and challenge fairness, enabling us to use identical patient cases for both tasks without disclosing Task 1's ground truth segmentation masks to Task 2 participants. To ensure practical implementation and efficient evaluation, we established specific technical constraints, namely that algorithms were required to complete processing within 20 minutes per patient case through the Grand Challenge runtime environment (using an NVIDIA T4 graphics processing unit). To assist participants, we provided detailed examples of Docker image containerization on our GitHub repository [28]. We launched two distinct phases for each task on August 15th, 2024: a preliminary development phase and a final test phase.

The preliminary development phase served as a "practice" round, allowing participants to debug their algorithms and familiarize themselves with the Docker submission framework. During this optional but highly recommended phase, teams could make up to five valid submissions. We used data from two patients not included in the training set, selecting straightforward cases with easily identifiable segmentation targets to facilitate the debugging process. The two patients selected for the preliminary phase were both human papillomavirus (HPV)-positive with large GTVp and GTVn targets, with one patient's images featuring fat suppression and the other's without, providing participants exposure to different MRI acquisition techniques commonly encountered in the dataset. Results from this phase were immediately displayed on the leaderboard but did not impact the final rankings.

The final test phase, which was composed of 50 cases, determined the official evaluation and ranking of participants' algorithms. In contrast to the development phase, each team was limited to a single valid submission. This restriction ensured a fair comparison of each team's best-performing algorithm. The test set for this phase was entirely separate from the development phase data, providing a true measure of the algorithms' performance on unseen cases.

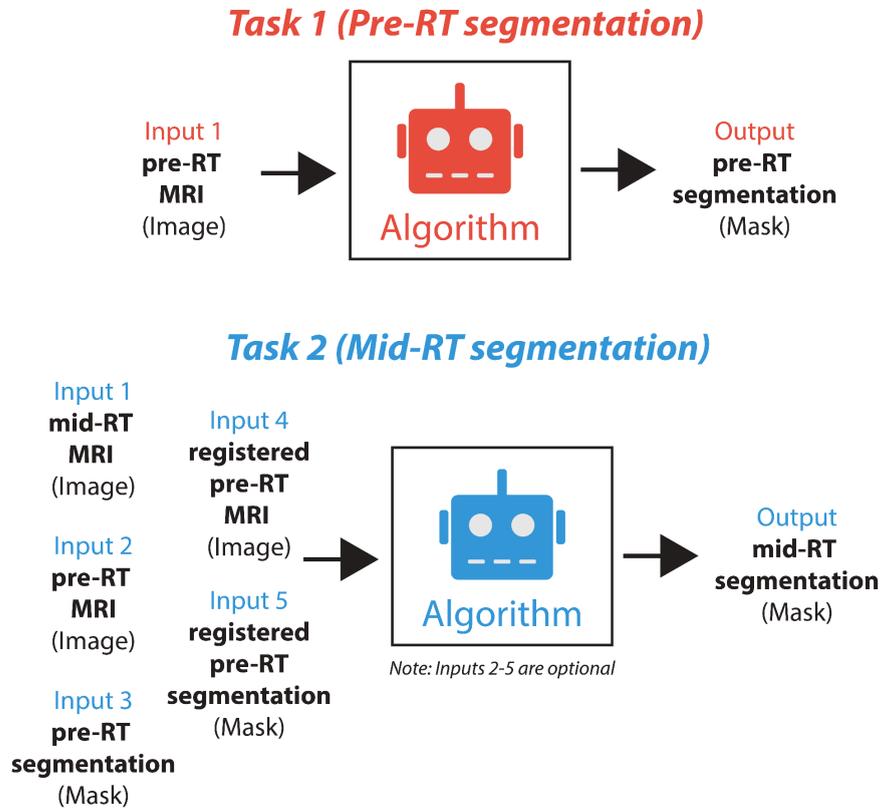

**Fig. 7.** Toy examples of model input and outputs for Task 1 (pre-radiotherapy segmentation, top) and Task 2 (mid-radiotherapy segmentation, bottom).

### 2.6    Baseline Models

To establish performance benchmarks for Tasks 1 and 2, we developed baseline algorithms using nnU-Net [33], widely regarded as the current DL gold standard for medical image segmentation [34]. For Task 1, we implemented an nnU-Net v2 model with default parameters (full 3D resolution, 1000 epochs, 5-fold cross-validation), using only the pre-RT training images (n=150) as input. We applied an identical nnU-Net approach for Task 2, but utilized mid-RT images (n=150) for training instead. No post-processing was applied to baseline models. Model training was performed on a Lamda workstation with 4 NVIDIA RTX A6000 graphics processing units. DL training took approximately 24 hours per model. Additionally, we created a simple "null" algorithm for Task 2, which uses unmodified pre-treatment segmentations as mid-RT predictions. This approach mimics a typical starting point for segmentation adjustments in routine clinical workflows.

### 2.7    Post-Challenge Publications and Conference

To be eligible for the final ranking and prizes, participants were required to submit a concise paper detailing their methods. Teams that participated in both tasks (pre-RT and mid-RT segmentation) had the option to submit either a single comprehensive paper or two separate papers describing their approaches. These submissions were subsequently published in our post-challenge proceedings, providing a valuable resource for the research community. Following the challenge's conclusion, we hosted a live virtual webinar event on the Zoom video conference platform, where top-performing teams were invited to present their innovative methods. This event culminated in the official announcement of the challenge winners.

## 3    Challenge Algorithm Results

### 3.1    Participation

As of September 18, 2024 (submission deadline), the number of registered teams for the challenge (regardless of the tasks) was 107. For each task, each team could submit up to five valid submissions for the preliminary development phase and one valid submission for the final test phase. By the submission deadline, we received a total of 164 valid entries across both tasks: 95 for Task 1 (75 in the preliminary development phase, 20 in the final test phase) and 69 for Task 2 (54 in the preliminary development phase, 15 in the final test phase). After accounting for eligibility, 19 unique teams were identified. The geographical distribution of initial registrants is shown in **Figure 8A**, while the distribution of final eligible participants is shown in **Figure 8B**. The geographical distribution of initial registrants and final participants followed similar patterns, except for Europe and Asia, where the relative proportions reversed between the initial registrants and the final eligible participants.

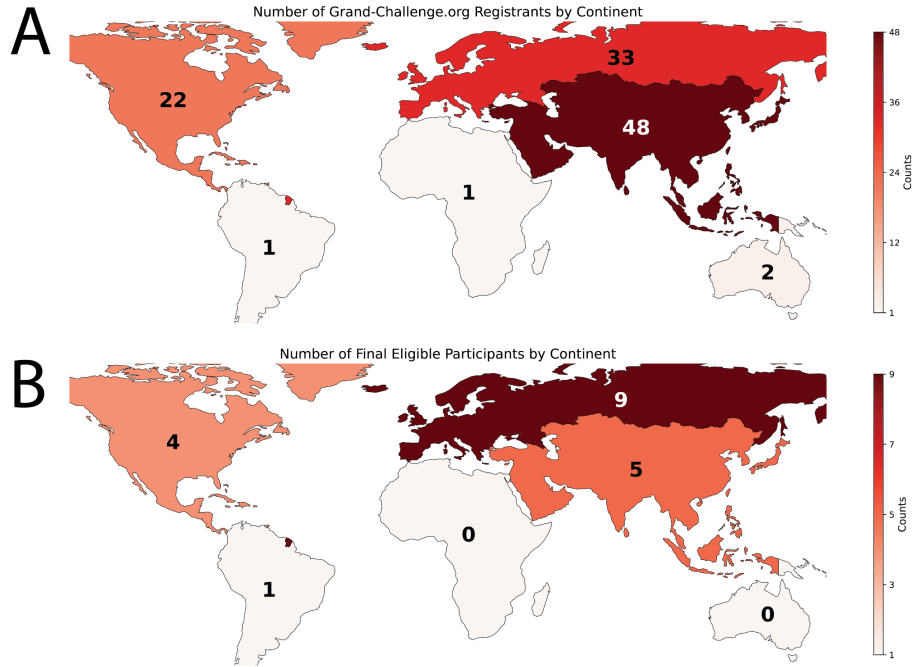

**Fig. 8.** Geographical distribution of initial registrants and final participants in the HNTS-MRG 2024 challenge by continent. (A) Number of initial registrants, showing the distribution of participants who signed up for the competition on the Grand Challenge website. (B) Number of final eligible participants, reflecting the participants who completed test phase submissions and submitted corresponding manuscripts. The color scale in both maps represents the count of individual Grand Challenge accounts per continent. For the final eligible participants, only the primary contact person from each team was considered.

### 3.2    Task 1 (Pre-RT Segmentation) Specific Results

**Summary of Participants Methods.**

This section provides an overview of the methods proposed by each team for the automatic segmentation of the GTVp and GTVn in Task 1. The descriptions are presented in the order of the official rankings, beginning with the top-performing team. Each method is briefly outlined, focusing on the key distinguishing features of the method and corresponding submitted manuscript.

Team TUMOR [35] experimented with various nnU-Net methodologies alongside MedNeXt transformer-based models [36] of different kernel sizes. Their models were pre-trained on mid-RT and registered pre-RT images, then fine-tuned on the original pre-RT images. They also explored various ensembling strategies but found that averaging nnU-Net and MedNeXt solutions resulted in worse performance than using either model individually. Ultimately, their best approach was a "small" MedNeXt model with a kernel size of 3, which was used for the final test phase submission of Task 1. Interestingly, they also experimented with fine-tuning using a public meningioma RT dataset [37] but found it did not enhance performance, likely due to discrepancies in features learned during pre-training.

Team Hilab [38] explored a fully supervised learning approach enhanced with pre-trained weights and data augmentation techniques. Notably, they used the SegRap2023 challenge dataset [9] for fully supervised pre-training and applied histogram matching during preprocessing to align intensity differences between CT and MRI data, along with nonlinear transformations to image intensities. To mitigate the impact of negative samples and encourage the network to learn class distributions more effectively, they employed the MixUp technique [39], which augments the dataset by creating new training examples through linear interpolation between sample pairs. For their Task 1 submission, they ultimately used a combination of their base model and  pre-training + MixUp model, incorporating ensembling from cross-validation folds.

Team Stockholm_Trio [40] explored a variety of architectures, including SegResNet, nnU-Net, ResEnc, MedNeXt, and U-Mamba. For Task 1, all models were trained exclusively on pre-RT data. The ResEnc and MedNeXt models consistently outperformed the others, particularly when trained on the preprocessed dataset (crop + intensity standardization). The best results were achieved by ensembling the ResEnc and MedNeXt models. However, the execution time of their Docker image on the evaluation platform exceeded the time limit, causing job submission failures. To meet the resource constraints, the final submitted models were significantly simplified by omitting certain  preprocessing and ensembling steps.

Team CEMRG [41] introduced a two-stage self-supervised learning approach which leverages unlabeled data to develop robust pre-trained models. In the first stage, they utilized a Self-Supervised Student-Teacher Learning Framework, specifically DINOv2 [42] adapted for 3D data, to learn effective representations from a limited unlabeled dataset. In the second stage, they fine-tuned an xLSTM-based [43] UNet model designed to capture both spatial and sequential features. For Task 1, the team fine-tuned their model on pre-RT segmentation data, which was ultimately submitted for the final test phase.

Team mic-dkfz [44] utilized nnU-net with a residual encoder architecture [34] for their Task 1 solution. Importantly, they experimented with various training strategies, including extensive data augmentation (Aug++), pretraining, ensembling, post-processing, and test-time augmentation. The team leveraged transfer learning by pretraining their model on an unprecedentedly large set of public 3D medical imaging datasets and then fine-tuning on pre-RT data. For their submission to the final test phase, they used an ensemble of models that combined Aug++  and pretraining.

Team RUG_UMCG [45] initially experimented with a custom framework for Task 1, incorporating MONAI [46] with U-Net, 3D U-Net, Swin UNETR architectures, and MedSAM — a foundation model pretrained on a large medical dataset [47]. However, these approaches were less optimal in terms of training speed, validation results, and overall performance compared to a vanilla nnU-net. Ultimately, they transitioned to the nnU-Net framework, employing a 15-fold cross-validation ensemble. For Task 1, they enhanced their training data by incorporating mid-RT data as separate inputs. During inference, test time augmentation was disabled to meet the challenge runtime limit.

Team alpinists [48] conducted a comprehensive literature review to inform their approach, ultimately proposing a resource-efficient two-stage segmentation method using nnU-Net with residual encoders. In this two-stage approach, the segmentation results from the first training round guided the sampling process for a second refinement stage. For the pre-RT task, they achieved competitive results using only the first-stage nnU-Net, which was submitted as their final model. For the final test set submission, they retrained their selected pre-RT models on the full 150-patient dataset. Uniquely, the team used Code-Carbon [49] to monitor the computational efficiency of their approach.

Team SZTU_SingularMatrix [50] investigated the use of STU-Net, a model designed to improve scalability and transferability for medical segmentation [51]. Their approach involved large-scale pretraining on datasets such as TotalSegmentator [52] followed by fine-tuning on the challenge dataset. They explored various STU-Net variants with different parameter sizes and ultimately selected the STU-Net-B model (featuring 58.26 million parameters) for their final test phase submission.

Team SJTU & Ninth People's Hospital [53] explored the use of an nnU-Net model with a residual encoder, coupled with explicit selection of training data. Their initial experiments showed poor performance when the model encountered cases with high background ratios. To address this, for their Task 1 approach they retrained the model using a carefully selected subset of data consisting of cases with background ratios of 70-90% (i.e., the proportion of background voxels to tumor voxels). This approach aimed to improve segmentation performance, and they also incorporated registered data in the training process. While the model performed well on cases with lower background ratios, further optimization was needed for cases with higher background ratios. For the Task 1 final test phase, the authors submitted an ensemble model trained specifically on cases with high background ratios.

Team DCPT-Stine [54] integrated two promising segmentation frameworks, UMamba [55] and nnU-Net with a residual encoder, into a new approach called UMambaAdj. This method combines the feature extraction strengths of the residual encoder with the long-range dependency capabilities of Mamba blocks. The proposed approach demonstrated comparable segmentation accuracy to the original UMambaEnc, but with reduced training and inference times. Additionally, the team found that UMamba blocks significantly improved distance-based metrics, although these metrics were not considered in the final challenge rankings.

Team UW LAIR [56] implemented SegResNet [57] with deep supervision for Task 1. They trained the model using both pre-RT and mid-RT data, but only pre-RT data was used for model selection in the validation set. For each training/validation split, they used three random seeds, selecting the model with the highest DSCagg in the validation set for each of the five cross-validation folds. This process was repeated twice with different random seeds, resulting in a total of 10 models. The final Task 1 submission was an ensemble of these 10 models.

Team NeuralRad [58] developed an enhanced nnU-Net model augmented with an autoencoder architecture. During inference, they added an output channel to predict the original input images, allowing the model to generate both segmentation results and autoencoder predictions simultaneously. By introducing the original training images as additional input channels and incorporating mean squared error loss alongside dice loss, the model was able to learn additional image features, improving segmentation accuracy.

Team 1WM [59] benchmarked several state-of-the-art segmentation architectures to determine whether recent advances in deep encoder-decoder models are effective for low-data and low-contrast tasks. Interestingly, their results showed that traditional UNet-based methods outperform more modern architectures like UNETR, SwinUNETR, and SegMamba, suggesting that factors like data preparation, the underlying objective function, and preprocessing play a greater role than the network architecture itself. For Task 1, they focused on a single-channel pre-RT network, using the pre-RT volume as input. Ultimately, they submitted a ResUNet 5-fold cross-validation ensemble model for the final test phase of Task 1.

Team dlabella29 [60] explored SegResNet integrated into Auto3DSeg via MONAI. The models were pre-trained on both pre-RT and mid-RT image-mask pairs and then fine-tuned on pre-RT data without any preprocessing. Extensive exploratory analysis of the training data also played a key role in shaping their post-processing decisions which included removing smaller tumor and node predictions. For their final test phase submission, they used an ensemble of six SegResNet models, fusing predictions through weighted majority voting.

Team PocketNet [61] implemented a lightweight CNN architecture called PocketNet [62] using the medical image segmentation toolkit [63]. Unlike traditional networks that double the number of feature maps at lower resolutions, PocketNet maintains a constant number of feature maps across all resolution levels. This design results in significantly faster training time while reducing memory usage and requirements. The PocketNet model was trained on pre-RT images via 5-fold cross validation and the corresponding ensemble was submitted for the final test phase of Task 1.

Team andrei.iantsen [64] explored different variations of standard U-Net architectures in their solutions. They tested various processing configurations, including normalization, augmentation, and weighting techniques to find an optimal approach. For Task 1, they trained their networks on all available MRI images, including pre-RT, mid-RT, and registered pre-RT images. In the final test phase of Task 1, they submitted a 5-fold cross-validation ensemble that combined patch-wise normalization, scheduled augmentation, and Gaussian weighting.

Team FinoxyAI [65] proposed a dual-stage 3D UNet approach, called DualUnet, which uses a cascaded Unet framework for progressive segmentation refinement. In the first stage, the models produce an initial binary segmentation, which is then refined by an ensemble of second-stage models to achieve multiclass segmentation. Both pre-RT and mid-RT MRI scans were used as training inputs. This dual-stage approach consistently outperformed single-stage methods in segmentation performance in validation experiments. The approach was trained using 5-fold cross-validation and submitted to the final test phase as an ensemble of five coarse models and ten refinement models.

Team ECU [66] employed LinkNet [67] ensembles for their solution. They initially pre-trained a LinkNet model with weights from ImageNet [68] followed by fine tuning on the challenge dataset. From the training process, they selected eight high-performing model weights to create an ensemble. Each selected weight was used to generate a LinkNet architecture, resulting in eight networks whose predictions were averaged to produce the final segmentation. Their validation experiments demonstrated that the ensemble outperformed any individual model. Interestingly, they also found that increasing the number of networks beyond eight did not significantly enhance accuracy, suggesting a point of diminishing returns. They suggest their approach leverages the benefits of ensemble learning without the computational cost of training each network from scratch.

**Challenge Ranking Results.**

The results for Task 1 are reported in **Table 2**. The DSCagg-mean results from the 18 participants ranged from 0.571 to 0.825 (overall mean = 0.783). Team TUMOR achieved the highest overall performance, with a DSCagg-mean of 0.825, including the top GTVn DSCagg score of 0.873. Team Stockholm_Trio secured the best GTVp DSCagg result, with a score of 0.795. Only the top 3 teams achieved DSCagg-mean results higher than the nnU-Net baseline (0.817). The top 9 teams (top 50%) achieved DSCagg-mean results higher than interobserver variability (0.806). Notably, two participants for Task 1 withdrew from the competition before submitting their methods manuscripts; their results are displayed for completeness in Table 2 but are not incorporated into any analysis.

**Table 2.** Task 1 (pre-radiotherapy segmentation) results for participating teams. Results are shown in descending order by mean aggregated DSC (DSCagg). DSCagg scores for primary gross tumor volume (GTVp) and metastatic lymph nodes (GTVn) are also shown. Highest scores for each category are bolded. The performance of the mean, baseline nnU-Net, interobserver variability (derived in section 2.3), anonymized withdrawn teams are also shown at the bottom of the table. Values are rounded to the nearest 3rd decimal place.

| Position | Team name | DSCagg mean | DSCagg GTVp | DSCagg GTVn |
|----------|-----------|-------------|-------------|-------------|
| 1 | TUMOR | **0.825** | 0.778 | **0.873** |
| 2 | HiLab | 0.824 | 0.785 | 0.862 |
| 3 | Stockholm_Trio | 0.822 | **0.795** | 0.849 |
| 4 | CEMRG | 0.815 | 0.769 | 0.860 |
| 5 | mic-dkfz | 0.812 | 0.756 | 0.869 |
| 6 | RUG_UMCG | 0.812 | 0.773 | 0.851 |
| 7 | alpinists | 0.810 | 0.767 | 0.852 |
| 8 | SZTU-SingularMatrix | 0.807 | 0.759 | 0.854 |

| 9 | SJTU&NINTH PEOPLE'S HOSPITAL | 0.806 | 0.767 | 0.845 |
|---|---|---|---|---|
| 10 | DCPT-Stine's group | 0.796 | 0.751 | 0.842 |
| 11 | UW LAIR | 0.794 | 0.745 | 0.844 |
| 12 | NeuralRad | 0.792 | 0.732 | 0.852 |
| 13 | IWM | 0.771 | 0.717 | 0.826 |
| 14 | dlabella29 | 0.771 | 0.720 | 0.822 |
| 15 | PocketNet | 0.770 | 0.732 | 0.808 |
| 16 | andrei.iantsen | 0.752 | 0.709 | 0.794 |
| 17 | FinoxyAI | 0.737 | 0.697 | 0.777 |
| 18 | ECU | 0.571 | 0.495 | 0.646 |
| NA | Mean | 0.783 | 0.734 | 0.829 |
| NA | nnU-Net baseline | 0.817 | 0.769 | 0.865 |
| NA | Interobserver variability | 0.806 | 0.757 | 0.854 |
| NA | Withdrawn team 1 | 0.774 | 0.726 | 0.822 |
| NA | Withdrawn team 2 | 0.793 | 0.751 | 0.836 |

### 3.3    Task 2 (Mid-RT Segmentation) Specific Results

**Summary of Participants Methods.**

This section provides an overview of the methods proposed by each team for the automatic segmentation of the GTVp and GTVn in Task 2. The descriptions are presented in the order of the official rankings, beginning with the top-performing team. Each method is briefly outlined, focusing on the key distinguishing features of the method and/or corresponding manuscript.

Team UW LAIR [56] integrated novel mask-aware attention modules into a SegResNet framework, allowing pre-RT masks to influence features learned from paired mid-RT data. The model took mid-RT MRI images along with pre-RT masks as inputs. During training, paired pre-RT data was also included, with prior masks set to zeros. They also applied mask propagation through deformable registration, which excluded predicted segmentations on mid-RT MRI scans that had no overlap with registered pre-RT segmentations. Ultimately, the attention-based approach outperformed the baseline method which concatenated mid-RT images with pre-RT masks. As in their Task 1 approach, they generated several models using cross validation splits and random seeds then ensembled them for the final submission.

Team mic-dkfz [44] integrated registered pre-RT images and their segmentations as additional inputs into the nnU-Net framework, through a method they referred to as LongiSeg [69]. Interestingly, though they initially experimented with the residual encoder architecture, it did not yield improved results. They investigated several LongiSeg variants and ultimately submitted an ensemble of their LongiSeg Pre-Seg-C model for the Task 2 final test phase. In this model, a one-hot encoding of the registered prior scan's segmentation mask was added to the network input (in addition to image inputs), following the order (current scan, prior scan, prior mask). The model was trained in chronological order, with the mid-RT scan as the current scan and the pre-RT scan as the prior.

Team HiLab [38] introduced an innovative training strategy with a novel network architecture, termed Dual Flow UNet, which features separate encoders for mid-RT images and registered pre-RT images along with their labels. In this setup, the mid-RT encoder progressively integrates information from pre-RT images and labels during forward propagation. Their submission to the Task 2 final test phase was an intricate ensemble of methods, combining folds from the Dual Flow UNet with base+pre-RT and pre-training + MixUp variants.

Team andrei.iantsen [64] used the same standard Unet based processing approaches for Task 2 as in Task 1. Notably, for Task 2 they trained models using four simultaneous input channels: the mid-RT image, registered pre-RT image, and two binary masks for the GTVp and GTVn on the registered pre-RT image. As in Task 1, their submitted 5-fold cross validation ensemble model for the Task 2 final test phase utilized all three modifications (patch-wise normalization, scheduled augmentation, gaussian weighting).

Team Stockholm_Trio [40] used the same architectures for Task 2 as in Task 1, with the addition of ablation studies to test different combinations of image data and segmentation masks. Due to difficulties in optimizing the MedNeXt model for this task, only the nnU-Net ResEnc model was ultimately used for training. Notably, they applied a dilation to the pre-RT masks, then derived signed distance maps from the dilated masks, incorporating them as prior information to guide the network's attention (preDistance-prior). Their results showed that the preDistance-prior settings outperformed other models. As with Task 1, to meet resource constraints, the final submitted models were simplified by omitting preprocessing and ensembling steps.

Team lWM [59] implemented similar experiments for Task 2 as in Task 1, this time using a three-channel mid-RT network with the concatenated mid-RT volume, registered pre-RT volume, and the associated registered pre-RT ground truth segmentation. As with Task 1, they found simple Unet models superior and subsequently submitted a ResUNet 5-fold-cross-validation ensemble model for the final test phase of Task 2.

Team RUG_UMCG [45] applied a similar approach to their Task 1 solution, utilizing the nnU-Net framework with a 15-fold cross-validation ensemble. For Task 2, they implemented a 3-channel input, where the mid-RT MRI volume served as the first channel, the registered pre-RT MRI volume as the second channel, and the corresponding segmentation mask as the third channel. As in Task 1, test-time augmentation was disabled during inference to comply with the challenge's runtime limit.

Team TUMOR [35] applied similar training approaches and architectures for Task 2 as in their Task 1 experiments, with the key difference being the use of concatenated multi-channel inputs to improve segmentation performance. Interestingly, they observed that using registered pre-RT images alone, without their segmentation masks, did not contribute useful information for segmenting mid-RT images. However, including both registered pre-RT images and their segmentation masks improved the DSCagg for mid-RT segmentation. Ultimately, an nnU-Net ensemble of the full-resolution and cascade models was selected as the final model for Task 2.

Team DCPT-Stine [70] employed a novel approach that computes gradient maps from pre-RT images and applies them to mid-RT images to enhance tumor boundary delineation. They applied connected component analysis to registered pre-RT tumor segmentations to initially create bounding boxes. These regions were then used to generate gradient maps on mid-RT T2w images, which served as additional input channels. Gradient maps from pre-RT images and their ground truth segmentation were also incorporated as extra training data. The method was built on nnU-Net with a residual encoder, and validation results showed that leveraging pre-RT information improved segmentation results. Experiments showed that using gradient maps led to more precise boundary localization than images alone.

Team alpinists [48] applied a similar two-stage approach as their Task 1 approach. However, to enhance segmentation performance, they incorporated prior knowledge from the registered pre-RT images and masks as an additional input for the second-stage refinement network. By leveraging the pre-RT data in the second stage, they were able to achieve more accurate mid-RT segmentations for their final submission. As with their Task 1 solution, they retrained the selected mid-RT model on the full 150-patient dataset.

Team NeuralRad [58] applied a similar approach for Task 2 as their Task 1 solution which coupled nnUnet to an autoencoder architecture. The main difference in their Task 2 submission was utilizing mid-RT data instead of pre-RT data.

Team dlabella29 [60] applied a similar methodology using SegResNet for Task 2 as they did in Task 1. Specific Task 2 preprocessing involved setting all voxels more than 1 cm from the registered pre-RT masks to background, followed by applying a bounding box to the image. The modified registered pre-RT and mid-RT MRI were used as input, and model training involved a single stage without any pre-training or fine tuning. Interestingly, they explored systematic radial reductions in the registered pre-RT masks and found that this simple technique performed surprisingly well. However, in keeping with the challenge's spirit, they avoided using simple mask reductions as their submission, as this method is not suitable for adaptive RT planning where patient-specific solutions are essential. Ultimately, they submitted an ensemble of five SegResNet models for the Task 2 final test phase submission.

Team CEMRG [41] employed the same two-stage self-supervised approach as in Task 1, but for Task 2, they fine-tuned their xLSTM-based UNet model on mid-RT segmentation data. This enabled the model to incorporate temporal dependencies specific to mid-treatment tumor response.

Team SJTU & Ninth People's Hospital [53] applied the same approach for Task 2 as they did for Task 1, using the nnU-Net residual encoder model coupled with selective training on specific data subsets. The key difference for Task 2 was the inclusion of mid-RT images instead of pre-RT images for model training. As in Task 1, they submitted an ensemble of folds for the final test phase.

Team TNL_skd [71] proposed an end-to-end coarse-to-fine cascade framework based on a 3D U-Net, inspired by future frame prediction in natural images and video [72]. The model has two interconnected components: a coarse segmentation network and a fine segmentation network, both sharing the same architecture. During coarse segmentation, a dilated pre-RT mask and mid-RT image are used to localize the region of interest and generate a preliminary prediction. During fine segmentation, resampling focuses on the region of interest, refining the prediction with the mid-RT image to produce the final mask. Notably, they also investigated training the networks separately but found the end-to-end combined model was superior.

**Challenge Ranking Results.**

The results for Task 2 are reported in **Table 3**. The DSCagg-mean results from the 15 participants ranged from 0.562 to 0.733 (overall mean = 0.688). Team UW LAIR achieved the highest overall performance, with a DSCagg-mean of 0.733, including the top GTVp DSCagg score of 0.607. Team mic-dkfz secured the best GTVn DSCagg result, with a score of 0.875. All teams, with the exception of one, achieved DSCagg-mean results higher than the nnU-Net baseline (0.633) and the null algorithm (0.601). Only the top 4 teams (~top 25%) achieved DSCagg-mean results higher than interobserver variability (0.714).

**Table 3.** Task 2 (mid-radiotherapy segmentation) results for participating teams. Results are shown in descending order by mean aggregated DSC (DSCagg). DSCagg scores for primary gross tumor volume (GTVp) and metastatic lymph nodes (GTVn) are also shown. Highest scores for each category are bolded. The performance of the mean, baseline nnU-Net, null algorithm (simple structure propagation from registered images), and interobserver variability (derived in section 2.3) are also shown at the bottom of the table. Values are rounded to the nearest 3rd decimal place.

| Position | Team name | DSCagg mean | DSCagg GTVp | DSCagg GTVn |
|---|---|---|---|---|
| 1 | UW LAIR | **0.733** | **0.607** | 0.859 |
| 2 | mic-dkfz | 0.727 | 0.579 | **0.875** |
| 3 | HiLab | 0.725 | 0.579 | 0.871 |
| 4 | andrei.iantsen | 0.718 | 0.592 | 0.845 |
| 5 | Stockholm_Trio | 0.710 | 0.554 | 0.866 |
| 6 | lWM | 0.707 | 0.579 | 0.836 |

| 7 | RUG_UMCG | 0.701 | 0.543 | 0.859 |
|---|---|---|---|---|
| 8 | TUMOR | 0.700 | 0.549 | 0.852 |
| 9 | DCPT-Stine's group | 0.700 | 0.534 | 0.867 |
| 10 | alpinists | 0.698 | 0.539 | 0.858 |
| 11 | NeuralRad | 0.685 | 0.526 | 0.843 |
| 12 | dlabella29 | 0.655 | 0.499 | 0.811 |
| 13 | CEMRG | 0.654 | 0.534 | 0.773 |
| 14 | SJTU&NINTH    PEOPLE'S HOSPITAL | 0.638 | 0.446 | 0.831 |
| 15 | TNL_skd | 0.562 | 0.500 | 0.625 |
| NA | Mean | 0.688 | 0.544 | 0.831 |
| NA | nnU-Net baseline | 0.633 | 0.422 | 0.844 |
| NA | null algorithm | 0.601 | 0.460 | 0.743 |
| NA | Interobserver variability | 0.714 | 0.600 | 0.828 |

## 3.4    General Results Summary

A boxplot summarizing both Task 1 and Task 2 performance is shown in **Figure 9**.

A correlation analysis was conducted to assess the relationship between participant performance in Task 1 and Task 2, including only those participants who completed both tasks. Correlations were generally weak with no significant relationships identified. Kendall's Tau correlation coefficients and corresponding p-values were: DSCagg-mean (-0.01, p=1.00), DSCagg-GTVp (0.01, p=1.00), and DSCagg-GTVn (0.18, p=0.38).

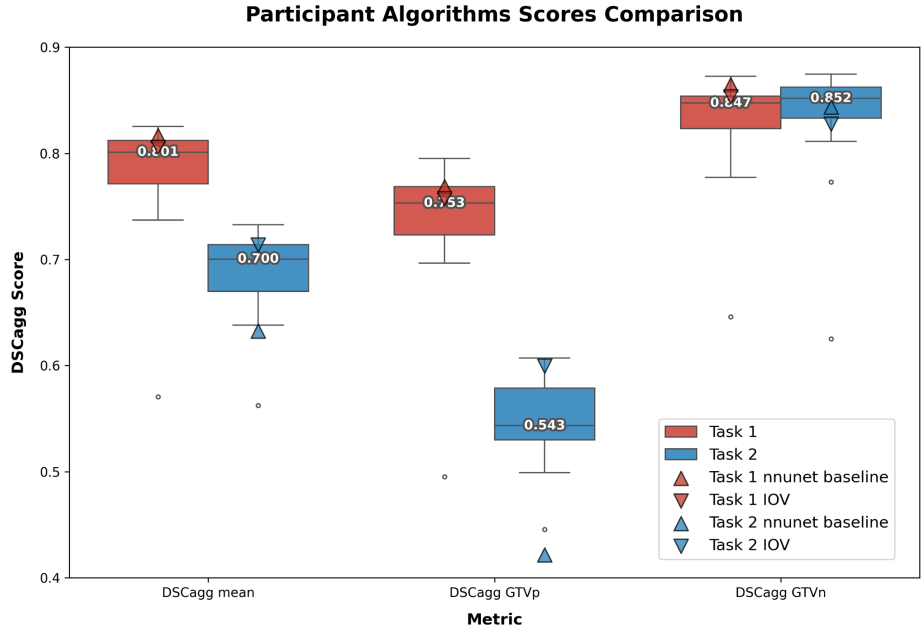

**Fig. 9.** Boxplot comparison of aggregated Dice Similarity Coefficient (DSCagg) scores across Task 1 (pre-radiotherapy) and Task 2 (mid-radiotherapy) for three metrics: DSCagg mean, DSCagg primary gross tumor volume (GTVp), and DSCagg metastatic lymph nodes (GTVn). Each box represents the interquartile range, with the horizontal line indicating the median score. Outliers are shown as individual points outside the whiskers. Task 1 is represented in red, and Task 2 in blue. Scatter symbols indicate the nnU-Net baseline (triangle), and interobserver variability (IOV, inverted triangle).

## 4      Discussion: Putting the Results into Context

### 4.1      Outcome and Findings

Data challenges play a crucial role in advancing research and facilitating the clinical implementation of AI technologies [73]. Our challenge represents the first crowdsourced initiative for MR-based segmentation in HNC, with a unique focus on investigating whether incorporating prior timepoint data enhances auto-segmentation performance in RT applications. This approach addresses a critical gap in the field and provides valuable insights for adaptive RT workflows.

Task 1 (pre-RT segmentation) results demonstrated the high performance of auto-segmentation algorithms, with most solutions predominantly based on nnU-Net architectures. It was shown that the top 50% of submitted methods achieved DSCagg-mean scores comparable to or exceeding our measured IOV (DSCagg-mean

~0.80), indicating their potential for clinical application. Our results align closely with those of HECKTOR 2022 [14], which also used DSCagg as a metric and saw top-performing algorithms achieve scores around 0.80, though their more heterogeneous test set potentially posed a more complex segmentation end-goal. Generally, GTVp structures were harder for algorithms to segment than GTVn structures. While teams experimented with various underlying training strategies and DL architectures, there didn't seem to be a clear optimal strategy for maximizing performance, though it is worth noting two out of the top three teams used MedNeXt — a transformer-driven architecture [36] — in their approach. Moreover, our baseline nnU-Net algorithm already achieved high-performing results, suggesting that current state-of-the-art methods provide a solid foundation for further improvements. The minimal quantitative differences observed between top-performing models echo findings from previous HNC challenges like HECKTOR [14], SegRap [9], and H&N-Seg [15]. This consistency across challenges underscores the robustness of current pre-RT segmentation algorithms, particularly given the strong baseline performance of nnU-Net. It suggests that for this specific task, we may be approaching a performance plateau with current DL architectures and available training data.

Task 2 (mid-RT segmentation) presented a more challenging problem, as clearly evidenced by the lower overall algorithmic performance compared to Task 1. This aligns with our expectations and the higher IOV observed in mid-RT annotations (DSCagg-mean ~ 0.71). As with Task 1, algorithms found GTVp structures more challenging to segment than GTVn structures. However, in Task 2, this difficulty gap was wider, aligning with the trends observed in our interobserver variability data. Notably, along with volumetric changes, tumor shrinkage is often accompanied by other radiation-induced biological effects [74], such as inflammation and necrosis. These changes can be visible on imaging and may complicate the accurate contouring of intra-treatment scans [75], particularly for GTVp structures. Subsequently, our baseline nnU-Net model for this task fell short of the DSCagg-mean IOV, likely due to the challenge of accurately capturing these complex, evolving tumor characteristics. Interestingly, a simple "null" model mimicking static contour propagation performed surprisingly similar to the baseline nnU-Net model (DSCagg-mean ~0.60). It's worth noting that our measured IOV may be slightly inflated due to the exclusion of particularly challenging cases (see **Section 2.3**, Sources of Errors - Interobserver Variability), potentially setting a higher benchmark than typically expected. Importantly, the vast majority of submitted algorithms thoroughly outperformed the baselines, with some even surpassing the IOV threshold. This achievement underscores the potential value of advanced auto-segmentation methods in adaptive RT workflows. However, the fact that only about 25% of teams were able to surpass IOV for Task 2, compared to 50% for Task 1, highlights the novel nature of this segmentation challenge and the need for innovative approaches. Moreover, GTVp IOV was only crossed by the winning algorithm, further illustrating the need to focus on GTVp auto-contouring improvements. Interestingly, the average GTVn segmentation performance was higher in Task 2 than in Task 1, likely because most OPC GTVn remain large and do not achieve a complete response by mid-RT [76], simplifying the segmentation process, especially if prior segmentation masks were incorporated. As expected, the most successful methods thoughtfully incorporated registered pre-RT data (i.e., images and masks) typically through novel DL architectural modifications, demonstrating the utility of leveraging prior timepoint information in adaptive RT auto-segmentation solutions. Although this challenge only utilized prior information from pre-RT to mid-RT scans, the same frameworks could potentially be extended to incorporate imaging data from additional intra-treatment timepoints.

## 4.2      Limitations of the Challenge

While we have striven for a comprehensive data challenge with adequate documentation, curation efforts, and execution, our study is not without extant limitations.

Firstly, a primary limitation of our data challenge was the relatively modest patient cohort size, derived from a single institutional data source. Our total cohort size (~ 200 cases) is on par with some previous HNC challenges like SegRap 2023 [9], but falls considerably short of larger-scale initiatives such as HECKTOR 2022 [14] (~ 900 cases). It is worth noting that, despite this limitation, DL auto-segmentation algorithms have demonstrated remarkable performance even with limited data [77, 78], as evidenced by the high performance achieved on our test sets. Nevertheless, expanding our dataset over time, following the example set by challenges like HECKTOR, would be beneficial for future iterations. On a related note, another limitation is our focus on oropharyngeal regional tumors, which restricts the diversity of HNC subsites represented in our study. While broadening the range of HNC regional subsites would be valuable, it's important to consider the potential advantages of a more focused approach. Recent recommendations in DL auto-segmentation suggest that decomposing tasks to reduce class imbalance (e.g., focusing on oropharyngeal region) may lead to more effective data utilization and superior models [79]. This data-centric approach could potentially yield better results than a single, all-encompassing model for diverse HNC subsites.

A second significant limitation was the high degree of IOV in our annotations, particularly evident in mid-RT GTVp structures. This variability, while expected, aligns with existing literature on human IOV in HNC tumor segmentation using MRI [80]. To address this issue in future challenges, it would be beneficial to implement strict annotation guidelines for clinician annotators. While such guidelines exist for clinical target volumes [81, 82], they are notably absent for gross tumor volumes, highlighting an area for improvement in future iterations. Furthermore, our study's reliance solely on MRI, while valuable for MR-centric adaptive approaches (e.g., MR-Linac), may have limited the accuracy of tumor delineation. Incorporating additional systematically co-registered imaging modalities such as PET and CT could enhance both the generation of ground truth segmentations by physicians and overall model performance, as supported by previous research [83, 84]. Although we initially planned to include multiple MRI sequences, particularly diffusion weighted sequences (i.e., apparent diffusion coefficient maps), data curation constraints prevented this inclusion without significantly reducing our sample size. Future challenges should explore the integration of additional MRI sequences (i.e., multiparametric MRI), as they could provide crucial information for more precise tumor segmentation [85].

Finally, while we aimed for a robust evaluation using DSCagg, a metric previously validated by Andrearczyk et al. [30, 32], our choice of evaluation metrics could be expanded in future iterations. Recent tumor segmentation challenges involving multiple objects, such as BraTS-METS 2024 [86], have employed more sophisticated measures like lesion-wise DSC. This approach uses ground truth label dilation to better understand lesion extent and rigorously penalizes false positives and negatives with a score of 0. Furthermore, related metrics developed for multiple sclerosis lesions, such as the object-normalized DSC proposed by Raina et al. [87], might offer advantages in handling volume discrepancies. This adaptation of DSC scales precision at a fixed recall rate, addressing bias related to the occurrence rate of the positive class in the ground truth. Notably, for RT-related tasks, incorporating surface distance measurements [88] or spatially accounting for healthy tissue proximity [89] could also provide a more comprehensive evaluation. Additionally, treating the nodal component of these tasks as an object detection problem in conjunction with segmentation could offer a more nuanced assessment of algorithm performance. In future iterations, adopting a broader range of evaluation metrics would likely provide a more holistic understanding of algorithm performance and better align with the specific intricacies of HNC segmentation for MRI-guided RT.

### 4.3    Future of the Challenge

While we initially released training data (i.e., MRI images and STAPLE consensus segmentations in NIfTI format) through Zenodo [31], we have plans for a more comprehensive data release. This expanded dataset will include raw DICOM data, individual observer segmentations, and relevant clinical metadata for both training and held-out evaluation sets. The inclusion of individual observer segmentations may be particularly valuable in ambiguity modeling experiments for deep learning uncertainty quantification [90, 91]. This extensive data release will be accompanied by a detailed data descriptor to facilitate its use by the research community. Furthermore, we intend to publish a post-challenge summary paper in a high-impact, field-specific journal. This paper will delve into meta-analytic approaches to comprehensively characterize algorithm results, including combined participant algorithms, inter-algorithm variability, additional subanalysis, and ranking stability, in a similar vein to previous post-challenge analyses [8, 32]. Eligible participants will be invited to co-author this manuscript, fostering collaborative insight into the challenge outcomes.

While there are currently no concrete plans for a second edition of HNTS-MRG, we remain open to the possibility of future iterations that could significantly enhance the challenge's scope and impact. Such future editions could potentially incorporate a wider array of imaging sequences and timepoints (i.e., greater number of intra-treatment images) [92], leveraging the full capabilities of MRI in adaptive RT for HNC. We also envision the inclusion of data from multiple institutions, which would not only increase the dataset size but also introduce valuable diversity in imaging protocols and patient populations for added generalization ability of algorithms. This approach would mirror the successful strategy employed by the HECKTOR series of challenges [14, 93, 94], which has seen progressive data enlargement and diversification over the years. By broadening our dataset in these ways, future iterations of HNTS-MRG could offer even more robust insights into the performance of MRI-guided RT segmentation algorithms.

# 5    Conclusions

This paper presented a comprehensive overview of the HNTS-MRG 2024 challenge, focusing on the automated analysis of MRI images in HNC patients. The challenge explored two critical tasks: fully-automated pre-RT segmentation (Task 1) and mid-RT segmentation (Task 2). Utilizing a robust dataset of 200 HNC cases (150 for training, 50 for final testing), this challenge garnered significant interest from leading research teams worldwide, resulting in 20 high-quality papers showcasing a diverse array of innovative methods. Task 1 algorithm performance was generally high and consistent with previous similar tumor segmentation challenges (e.g., HECKTOR 2022). Top-performing algorithms for Task 1 achieved DSCagg-mean scores comparable to or exceeding clinician IOV, with minimal differences between leading methods. Task 2 proved more challenging, as expected, with lower model performance compared to Task 1. Notably, the best-performing algorithms in Task 2 surpassed both our baseline models and clinician IOV, demonstrating the potential of advanced auto-segmentation methods in adaptive RT workflows. Across both tasks, algorithms consistently found GTVp structures more difficult to segment than GTVn structures, mirroring trends in clinician IOV. To further advance this field, future work should focus on harmonizing tumor segmentation guidelines for clinicians, investigating additional segmentation performance metrics, and expanding the patient cohort.

**Acknowledgments.** The organizing committee extends its sincere gratitude to all participating teams for their dedication, innovative approaches, and significant contributions to this challenge. This data challenge was directly supported by National Institutes of Health (NIH) Administrative Supplements to Support Collaborations to Improve the AI/ML-Readiness of NIH-Supported Data provided by the National Institute of Dental and Craniofacial Research (NIDCR)  (R01DE028290-05S1) and National Cancer Institute (NCI) (R01CA257814-03S2) under parent grants R01DE028290 and R01CA257814, respectively. K.A.W. and B.A.M. are supported by the NCI through Image Guided Cancer Therapy (IGCT) T32 Training Program Fellowships (T32CA261856). L.M. is supported by a NIH Diversity Supplement (R01CA257814-02S2). M.A.N. receives funding from NIH NIDCR Grant (R03DE033550). C.D.F. receives related support from the NCI MD Anderson Cancer Center Core Support Grant Image-Driven Biologically-informed Therapy (IDBT) Program (P30CA016672-47). The authors also acknowledge support from the Tumor Measurement Initiative through the MD Anderson Strategic Research Initiative Development Program.

# References

1. Pollard, J.M., Wen, Z., Sadagopan, R., Wang, J., Ibbott, G.S.: The future of image-guided radiotherapy will be MR guided. Br. J. Radiol. 90, 20160667 (2017).
2. Mulder, S.L., Heukelom, J., McDonald, B.A., Van Dijk, L., Wahid, K.A., Sanders, K., Salzillo, T.C., Hemmati, M., Schaefer, A., Fuller, C.D.: MR-guided adaptive radiotherapy for OAR sparing in head and neck cancers. Cancers . 14, 1909 (2022).
3. Salzillo, T.C., Taku, N., Wahid, K.A., McDonald, B.A., Wang, J., van Dijk, L.V., Rigert, J.M., Mohamed, A.S.R., Wang, J., Lai, S.Y., Fuller, C.D.: Advances in imaging for HPV-related oropharyngeal cancer: Applications to radiation oncology. Semin. Radiat. Oncol. 31, 371–388 (2021).
4. Kiser, K.J., Smith, B.D., Wang, J., Fuller, C.D.: "Après Mois, Le Déluge": Preparing for the Coming Data Flood in the MRI-Guided Radiotherapy Era. Front. Oncol. 9, 983 (2019).
5. Thorwarth, D., Low, D.A.: Technical Challenges of Real-Time Adaptive MR-Guided Radiotherapy. Front. Oncol. 11, 634507 (2021).
6. Segedin, B., Petric, P.: Uncertainties in target volume delineation in radiotherapy - are they relevant and what can we do about them? Radiol. Oncol. 50, 254–262 (2016).
7. Hindocha, S., Zucker, K., Jena, R., Banfill, K., Mackay, K., Price, G., Pudney, D., Wang, J., Taylor, A.: Artificial Intelligence for Radiotherapy Auto-Contouring: Current Use, Perceptions of and Barriers to Implementation. Clin. Oncol. . 35, 219–226 (2023).
8. Oreiller, V., Andrearczyk, V., Jreige, M., Boughdad, S., Elhalawani, H., Castelli, J., Vallières, M., Zhu, S., Xie, J., Peng, Y., Iantsen, A., Hatt, M., Yuan, Y., Ma, J., Yang, X., Rao, C., Pai, S., Ghimire, K., Feng, X., Naser, M.A., Fuller, C.D., Yousefirizi, F., Rahmim, A., Chen, H., Wang, L., Prior, J.O., Depeursinge, A.: Head and neck tumor segmentation in PET/CT: The HECKTOR challenge. Med. Image Anal. 77, 102336 (2022).
9. Luo, X., Fu, J., Zhong, Y., Liu, S., Han, B., Astaraki, M., Bendazzoli, S., Toma-Dasu, I., Ye, Y., Chen, Z., Xia, Y., Su, Y., Ye, J., He, J., Xing, Z., Wang, H., Zhu, L., Yang, K., Fang, X., Wang, Z., Lee, C.W., Park, S.J., Chun, J., Ulrich, C., Maier-Hein, K.H., Ndipenoch, N., Miron, A., Li, Y., Zhang, Y., Chen, Y., Bai, L., Huang, J., An, C., Wang, L., Huang, K., Gu, Y., Zhou, T., Zhou, M., Zhang, S., Liao, W., Wang, G., Zhang, S.: SegRap2023: A Benchmark of Organs-at-Risk and Gross Tumor Volume Segmentation for Radiotherapy Planning of Nasopharyngeal Carcinoma, http://arxiv.org/abs/2312.09576, (2023). https://doi.org/10.48550/ARXIV.2312.09576.
10. Maier-Hein, L., Reinke, A., Kozubek, M., Martel, A.L., Arbel, T., Eisenmann, M., Hanbury, A., Jannin, P., Müller, H., Onogur, S., Saez-Rodriguez, J., van Ginneken, B., Kopp-Schneider, A., Landman, B.A.: BIAS: Transparent reporting of biomedical image analysis challenges. Med. Image Anal. 66, 101796 (2020).
11. Pinkiewicz, M., Dorobisz, K., Zatoński, T.: A Systematic Review of Cancer of Unknown Primary in the Head and Neck Region. Cancer Manag. Res. 13, 7235–7241 (2021).
12. Burnet, N.G., Thomas, S.J., Burton, K.E., Jefferies, S.J.: Defining the tumour and target volumes for radiotherapy. Cancer Imaging. 4, 153–161 (2004).
13. Jensen, K., Al-Farra, G., Dejanovic, D., Eriksen, J.G., Loft, A., Hansen, C.R., Pameijer,

F.A., Zukauskaite, R., Grau, C.: Imaging for Target Delineation in Head and Neck Cancer Radiotherapy. Semin. Nucl. Med. 51, 59–67 (2021).

14. Andrearczyk, V., Oreiller, V., Abobakr, M., Akhavanallaf, A., Balermpas, P., Boughdad, S., Capriotti, L., Castelli, J., Cheze Le Rest, C., Decazes, P., Correia, R., El-Habashy, D., Elhalawani, H., Fuller, C.D., Jreige, M., Khamis, Y., La Greca, A., Mohamed, A., Naser, M., Prior, J.O., Ruan, S., Tanadini-Lang, S., Tankyevych, O., Salimi, Y., Vallières, M., Vera, P., Visvikis, D., Wahid, K., Zaidi, H., Hatt, M., Depeursinge, A.: Overview of the HECKTOR Challenge at MICCAI 2022: Automatic Head and Neck Tumor Segmentation and Outcome Prediction in PET/CT. In: Head and Neck Tumor Segmentation and Outcome Prediction. pp. 1–30. Springer Nature Switzerland (2023).

15. Podobnik, G., Ibragimov, B., Tappeiner, E., Lee, C., Kim, J.S., Mesbah, Z., Modzelewski, R., Ma, Y., Yang, F., Rudecki, M., Wodziński, M., Peterlin, P., Strojan, P., Vrtovec, T.: HaN-Seg: The head and neck organ-at-risk CT and MR segmentation challenge. Radiother. Oncol. 198, 110410 (2024).

16. Boettiger, C.: An introduction to Docker for reproducible research. Oper. Syst. Rev. 49, 71–79 (2015).

17. Head and neck tumor segmentation for MR-guided applications - grand challenge, https://hntsmrg24.grand-challenge.org/, last accessed 2024/08/12.

18. McDonald, B.A., Dal Bello, R., Fuller, C.D., Balermpas, P.: The Use of MR-Guided Radiation Therapy for Head and Neck Cancer and Recommended Reporting Guidance. Semin. Radiat. Oncol. 34, 69–83 (2024).

19. Lin, D., Wahid, K.A., Nelms, B.E., He, R., Naser, M.A., Duke, S., Sherer, M.V., Christodouleas, J.P., Mohamed, A.S.R., Cislo, M., Murphy, J.D., Fuller, C.D., Gillespie, E.F.: E pluribus unum: prospective acceptability benchmarking from the Contouring Collaborative for Consensus in Radiation Oncology crowdsourced initiative for multiobserver segmentation. J Med Imaging (Bellingham). 10, S11903 (2023).

20. Wahid, K.A., Lin, D., Sahin, O., Cislo, M., Nelms, B.E., He, R., Naser, M.A., Duke, S., Sherer, M.V., Christodouleas, J.P., Mohamed, A.S.R., Murphy, J.D., Fuller, C.D., Gillespie, E.F.: Large scale crowdsourced radiotherapy segmentations across a variety of cancer anatomic sites. Sci. Data. 10, 161 (2023).

21. Warfield, S.K., Zou, K.H., Wells, W.M.: Simultaneous truth and performance level estimation (STAPLE): an algorithm for the validation of image segmentation. IEEE Trans. Med. Imaging. 23, 903–921 (2004).

22. Mayo, C.S., Moran, J.M., Bosch, W., Xiao, Y., McNutt, T., Popple, R., Michalski, J., Feng, M., Marks, L.B., Fuller, C.D., Yorke, E., Palta, J., Gabriel, P.E., Molineu, A., Matuszak, M.M., Covington, E., Masi, K., Richardson, S.L., Ritter, T., Morgas, T., Flampouri, S., Santanam, L., Moore, J.A., Purdie, T.G., Miller, R.C., Hurkmans, C., Adams, J., Jackie Wu, Q.-R., Fox, C.J., Siochi, R.A., Brown, N.L., Verbakel, W., Archambault, Y., Chmura, S.J., Dekker, A.L., Eagle, D.G., Fitzgerald, T.J., Hong, T., Kapoor, R., Lansing, B., Jolly, S., Napolitano, M.E., Percy, J., Rose, M.S., Siddiqui, S., Schadt, C., Simon, W.E., Straube, W.L., St James, S.T., Ulin, K., Yom, S.S., Yock, T.I.: American Association of Physicists in Medicine Task Group 263: Standardizing Nomenclatures in Radiation Oncology. Int. J. Radiat. Oncol. Biol. Phys. 100, 1057–1066 (2018).

23. Lowekamp, B.C., Chen, D.T., Ibáñez, L., Blezek, D.: The Design of SimpleITK. Front. Neuroinform. 7, 45 (2013).

24. Anderson, B.M., Wahid, K.A., Brock, K.K.: Simple Python Module for Conversions Between DICOM Images and Radiation Therapy Structures, Masks, and Prediction Arrays. Pract. Radiat. Oncol. 11, 226–229 (2021).

25. Wahid, K.A., Glerean, E., Sahlsten, J., Jaskari, J., Kaski, K., Naser, M.A., He, R., Mohamed, A.S.R., Fuller, C.D.: Artificial Intelligence for Radiation Oncology Applications Using Public Datasets. Semin. Radiat. Oncol. 32, 400–414 (2022).

26. Leibfarth, S., Mönnich, D., Welz, S., Siegel, C., Schwenzer, N., Schmidt, H., Zips, D.,

Thorwarth, D.: A strategy for multimodal deformable image registration to integrate PET/MR into radiotherapy treatment planning. Acta Oncol. 52, 1353–1359 (2013).

27. Naser, M.A., Wahid, K.A., Ahmed, S., Salama, V., Dede, C., Edwards, B.W., Lin, R., McDonald, B., Salzillo, T.C., He, R., Ding, Y., Abdelaal, M.A., Thill, D., O'Connell, N., Willcut, V., Christodouleas, J.P., Lai, S.Y., Fuller, C.D., Mohamed, A.S.R.: Quality assurance assessment of intra-acquisition diffusion-weighted and T2-weighted magnetic resonance imaging registration and contour propagation for head and neck cancer radiotherapy. Med. Phys. 50, 2089–2099 (2023).

28. Wahid, K.: HNTSMRG_2024: Docker tutorial and example files related to HNTS-MRG 2024 Data Challenge. Github.

29. Nikolov, S., Blackwell, S., Zverovitch, A., Mendes, R., Livne, M., De Fauw, J., Patel, Y., Meyer, C., Askham, H., Romera-Paredes, B., Kelly, C., Karthikesalingam, A., Chu, C., Carnell, D., Boon, C., D'Souza, D., Moinuddin, S.A., Garie, B., McQuinlan, Y., Ireland, S., Hampton, K., Fuller, K., Montgomery, H., Rees, G., Suleyman, M., Back, T., Hughes, C.O., Ledsam, J.R., Ronneberger, O.: Clinically Applicable Segmentation of Head and Neck Anatomy for Radiotherapy: Deep Learning Algorithm Development and Validation Study. J. Med. Internet Res. 23, e26151 (2021).

30. Andrearczyk, V., Oreiller, V., Jreige, M., Castelli, J., Prior, J.O., Depeursinge, A.: Segmentation and Classification of Head and Neck Nodal Metastases and Primary Tumors in PET/CT. Conf. Proc. IEEE Eng. Med. Biol. Soc. 2022, 4731–4735 (2022).

31. Wahid, K., Dede, C., Naser, M., Fuller, C.: Training Dataset for HNTSMRG 2024 Challenge, https://zenodo.org/doi/10.5281/zenodo.11199559, (2024). https://doi.org/10.5281/ZENODO.11199559.

32. Andrearczyk, V., Oreiller, V., Boughdad, S., Le Rest, C.C., Tankyevych, O., Elhalawani, H., Jreige, M., Prior, J.O., Vallières, M., Visvikis, D., Hatt, M., Depeursinge, A.: Automatic Head and Neck Tumor segmentation and outcome prediction relying on FDG-PET/CT images: Findings from the second edition of the HECKTOR challenge. Med. Image Anal. 90, 102972 (2023).

33. Isensee, F., Jaeger, P.F., Kohl, S.A.A., Petersen, J., Maier-Hein, K.H.: nnU-Net: a self-configuring method for deep learning-based biomedical image segmentation. Nat. Methods. 18, 203–211 (2021).

34. Isensee, F., Wald, T., Ulrich, C., Baumgartner, M., Roy, S., Maier-Hein, K., Jaeger, P.F.: nnU-Net Revisited: A Call for Rigorous Validation in 3D Medical Image Segmentation, http://arxiv.org/abs/2404.09556, (2024).

35. Moradi, N., Ferreira, A., Puladi, B., Kleesiek, J., Fatemizadeh, E., Luijten, G., Egger, J.: Comparative Analysis of nnUNet and MedNeXt for Head and Neck Tumor Segmentation in MRI-guided Radiotherapy. In: Wahid, K.A., Dede, C., Naser, M.A., and Fuller, C.D. (eds.) LNCS. Springer, Cham (2025).

36. Roy, S., Koehler, G., Ulrich, C., Baumgartner, M., Petersen, J., Isensee, F., Jaeger, P.F., Maier-Hein, K.: MedNeXt: Transformer-driven scaling of ConvNets for medical image segmentation, http://arxiv.org/abs/2303.09975, (2023).

37. LaBella, D., Khanna, O., McBurney-Lin, S., Mclean, R., Nedelec, P., Rashid, A.S., Tahon, N.H., Altes, T., Baid, U., Bhalerao, R., Dhemesh, Y., Floyd, S., Godfrey, D., Hilal, F., Janas, A., Kazerooni, A., Kent, C., Kirkpatrick, J., Kofler, F., Leu, K., Maleki, N., Menze, B., Pajot, M., Reitman, Z.J., Rudie, J.D., Saluja, R., Velichko, Y., Wang, C., Warman, P.I., Sollmann, N., Diffley, D., Nandolia, K.K., Warren, D.I., Hussain, A., Fehringer, J.P., Bronstein, Y., Deptula, L., Stein, E.G., Taherzadeh, M., Portela de Oliveira, E., Haughey, A., Kontzialis, M., Saba, L., Turner, B., Brüßeler, M.M.T., Ansari, S., Gkampenis, A., Weiss, D.M., Mansour, A., Shawali, I.H., Yordanov, N., Stein, J.M., Hourani, R., Moshebah, M.Y., Abouelatta, A.M., Rizvi, T., Willms, K., Martin, D.C., Okar, A., D'Anna, G., Taha, A., Sharifi, Y., Faghani, S., Kite, D., Pinho, M., Haider, M.A., Alonso-Basanta, M., Villanueva-Meyer, J., Rauschecker, A.M., Nada, A., Aboian, M.,

Flanders, A., Bakas, S., Calabrese, E.: A multi-institutional meningioma MRI dataset for automated multi-sequence image segmentation. Sci. Data. 11, 496 (2024).

38. Wang, L., Liao, W., Zhang, S., Wang, G.: Head and Neck Tumor Segmentation of MRI from Pre- and Mid-radiotherapy with Pre-training, Data Augmentation and Dual Flow UNet. In: Wahid, K.A., Dede, C., Naser, M.A., and Fuller, C.D. (eds.) LNCS. Springer, Cham (2025).

39. Zhang, H., Cisse, M., Dauphin, Y.N., Lopez-Paz, D.: mixup: Beyond Empirical Risk Minimization, http://arxiv.org/abs/1710.09412, (2017).

40. Astaraki, M., Toma-Dasu, I.: Enhancing Head and Neck Tumor Segmentation in MRI: The Impact of Image Preprocessing and Model Ensembling. In: Wahid, K.A., Dede, C., Naser, M.A., and Fuller, C.D. (eds.) LNCS. Springer, Cham (2025).

41. Qayyum, A., Mazher, M., Niederer, S.A.: Assessing Self-Supervised xLSTM-UNet Architectures for Head and Neck Tumor Segmentation in MR-Guided Applications. In: Wahid, K.A., Dede, C., Naser, M.A., and Fuller, C.D. (eds.) LNCS. Springer, Cham (2025).

42. Oquab, M., Darcet, T., Moutakanni, T., Vo, H., Szafraniec, M., Khalidov, V., Fernandez, P., Haziza, D., Massa, F., El-Nouby, A., Assran, M., Ballas, N., Galuba, W., Howes, R., Huang, P.-Y., Li, S.-W., Misra, I., Rabbat, M., Sharma, V., Synnaeve, G., Xu, H., Jegou, H., Mairal, J., Labatut, P., Joulin, A., Bojanowski, P.: DINOv2: Learning robust visual features without supervision, http://arxiv.org/abs/2304.07193, (2023).

43. Alkin, B., Beck, M., Pöppel, K., Hochreiter, S., Brandstetter, J.: Vision-LSTM: xLSTM as Generic Vision Backbone, http://arxiv.org/abs/2406.04303, (2024).

44. Kächele, J., Zenk, M., Rokuss, M., Ulrich, C., Wald, T., Maier-Hein, K.H.: Enhanced nnU-Net Architectures for Automated MRI Segmentation of Head and Neck Tumors in Adaptive Radiation Therapy. In: Wahid, K.A., Dede, C., Naser, M.A., and Fuller, C.D. (eds.) LNCS. Springer, Cham (2025).

45. Mol, F.N., van der Hoek, L., Ma, B., Nagam, B.C., Sijtsema, N.M., van Dijk, L.V., Bunte, K., Vlijm, R., van Ooijen, P.M.A.: MRI-based Head and Neck Tumor Segmentation Using nnU-Net with 15-fold Cross-Validation Ensemble. In: Wahid, K.A., Dede, C., Naser, M.A., and Fuller, C.D. (eds.) LNCS. Springer, Cham (2025).

46. Cardoso, M.J., Li, W., Brown, R., Ma, N., Kerfoot, E., Wang, Y., Murrey, B., Myronenko, A., Zhao, C., Yang, D., Nath, V., He, Y., Xu, Z., Hatamizadeh, A., Myronenko, A., Zhu, W., Liu, Y., Zheng, M., Tang, Y., Yang, I., Zephyr, M., Hashemian, B., Alle, S., Darestani, M.Z., Budd, C., Modat, M., Vercauteren, T., Wang, G., Li, Y., Hu, Y., Fu, Y., Gorman, B., Johnson, H., Genereaux, B., Erdal, B.S., Gupta, V., Diaz-Pinto, A., Dourson, A., Maier-Hein, L., Jaeger, P.F., Baumgartner, M., Kalpathy-Cramer, J., Flores, M., Kirby, J., Cooper, L.A.D., Roth, H.R., Xu, D., Bericat, D., Floca, R., Zhou, S.K., Shuaib, H., Farahani, K., Maier-Hein, K.H., Aylward, S., Dogra, P., Ourselin, S., Feng, A.: MONAI: An open-source framework for deep learning in healthcare, http://arxiv.org/abs/2211.02701, (2022).

47. Ma, J., He, Y., Li, F., Han, L., You, C., Wang, B.: Segment anything in medical images. Nat. Commun. 15, 654 (2024).

48. Tappeiner, E., Gapp, C., Welk, M., Schubert, R.: Head and Neck Tumor Segmentation on MRIs with Fast and Resource-Efficient Staged nnU-Nets. In: Wahid, K.A., Dede, C., Naser, M.A., and Fuller, C.D. (eds.) LNCS. Springer, Cham (2025).

49. Courty, B., Schmidt, V., Goyal-Kamal, MarionCoutarel, Blanche, L., Feld, B., inimaz, Lecourt, J., LiamConnell, SabAmine, supatomic, Lloret, P., Léval, M., Cruveiller, A., Saboni, A., ouminasara, Zhao, F., Joshi, A., Bauer, C., Bogroff, A., de Lavoreille, H., Laskaris, N., Phiev, A., Abati, E., rosekelly, Blank, D., Wang, Z., Otávio, L., Catovic, A.: mlco2/codecarbon: v2.8.0. Zenodo (2024). https://doi.org/10.5281/ZENODO.14212766.

50. Wang, Z., Lyu, M.: Head and Neck Tumor Segmentation for MRI-Guided Radiation Therapy Using Pre-trained STU-Net Models. In: Wahid, K.A., Dede, C., Naser, M.A., and

Fuller, C.D. (eds.) LNCS. Springer, Cham (2025).

51. Huang, Z., Wang, H., Deng, Z., Ye, J., Su, Y., Sun, H., He, J., Gu, Y., Gu, L., Zhang, S., Qiao, Y.: STU-Net: Scalable and Transferable medical image segmentation models empowered by large-scale supervised pre-training, http://arxiv.org/abs/2304.06716, (2023).

52. Wasserthal, J., Breit, H.-C., Meyer, M.T., Pradella, M., Hinck, D., Sauter, A.W., Heye, T., Boll, D.T., Cyriac, J., Yang, S., Bach, M., Segeroth, M.: TotalSegmentator: Robust segmentation of 104 anatomic structures in CT images. Radiol. Artif. Intell. 5, e230024 (2023).

53. Ji, K., Wu, Z., Han, J., Jia, J., Zhai, G., Liu, J.: Application of 3D nnU-Net with Residual Encoder in the 2024 MICCAI Head and Neck Tumor Segmentation Challenge. In: Wahid, K.A., Dede, C., Naser, M.A., and Fuller, C.D. (eds.) LNCS. Springer, Cham (2025).

54. Ren, J., Hochreuter, K., Kallehauge, J.F., Korreman, S.: UMambaAdj: Advancing GTV Segmentation for Head and Neck Cancer in MRI-Guided RT with UMamba and nnU-Net ResEnc Planner. In: Wahid, K.A., Dede, C., Naser, M.A., and Fuller, C.D. (eds.) LNCS. Springer, Cham (2025).

55. Ma, J., Li, F., Wang, B.: U-mamba: Enhancing long-range dependency for biomedical image segmentation, http://arxiv.org/abs/2401.04722, (2024).

56. Tie, X., Chen, W., Huemann, Z., Schott, B., Liu, N., Bradshaw, T.J.: Deep Learning for Longitudinal Gross Tumor Volume Segmentation in MRI-Guided Adaptive Radiotherapy for Head and Neck Cancer. In: Wahid, K.A., Dede, C., Naser, M.A., and Fuller, C.D. (eds.) LNCS. Springer, Cham (2025).

57. Myronenko, A.: 3D MRI brain tumor segmentation using autoencoder regularization, http://arxiv.org/abs/1810.11654, (2018).

58. An, Y., Wang, Z., Ma, E., Jiang, H., Lu, W.: Enhancing nnUNetv2 Training with Autoencoder Architecture for Improved Medical Image Segmentation. In: Wahid, K.A., Dede, C., Naser, M.A., and Fuller, C.D. (eds.) LNCS. Springer, Cham (2025).

59. Wodzinski, M.: Benchmark of Deep Encoder-Decoder Architectures for Head and Neck Tumor Segmentation in Magnetic Resonance Images: Contribution to the HNTSMRG Challenge. In: Wahid, K.A., Dede, C., Naser, M.A., and Fuller, C.D. (eds.) LNCS. Springer, Cham (2025).

60. LaBella, D.: Ensemble Deep Learning Models for Automated Segmentation of Tumor and Lymph Node Volumes in Head and Neck Cancer Using Pre- and Mid- Treatment MRI: Application of Auto3DSeg and SegResNet. In: Wahid, K.A., Dede, C., Naser, M.A., and Fuller, C.D. (eds.) LNCS. Springer, Cham (2025).

61. Twam, A., Celaya, A., Lim, E., Elsayes, K., Fuentes, D., Netherton, T.: Head and Neck Gross Tumor Volume Automatic Segmentation using PocketNet. In: Wahid, K.A., Dede, C., Naser, M.A., and Fuller, C.D. (eds.) LNCS. Springer, Cham (2025).

62. Celaya, A., Actor, J.A., Muthusivarajan, R., Gates, E., Chung, C., Schellingerhout, D., Riviere, B., Fuentes, D.: PocketNet: A smaller neural network for medical image analysis. IEEE Trans. Med. Imaging. 42, 1172–1184 (2023).

63. Celaya, A., Lim, E., Glenn, R., Mi, B., Balsells, A., Schellingerhout, D., Netherton, T., Chung, C., Riviere, B., Fuentes, D.: MIST: A simple and scalable end-to-end 3D medical imaging segmentation framework, http://arxiv.org/abs/2407.21343, (2024).

64. Iantsen, A.: Improving the U-Net Configuration for Automated Delineation of Head and Neck Cancer on MRI. In: Wahid, K.A., Dede, C., Naser, M.A., and Fuller, C.D. (eds.) LNCS. Springer, Cham (2025).

65. Saukkoriipi, M., Sahlsten, J., Jaskari, J., Al-Tahmeesschi, A., Ruotsalainen, L., Kaski, K.: Head and Neck Tumor Segmentation using Pre-RT MRI Scans and Cascaded DualUNet. In: Wahid, K.A., Dede, C., Naser, M.A., and Fuller, C.D. (eds.) LNCS. Springer, Cham (2025).

66. Baldeon-Calisto, M.: Ensemble of LinkNet networks for head and neck tumor

segmentation. In: Wahid, K.A., Dede, C., Naser, M.A., and Fuller, C.D. (eds.) LNCS. Springer, Cham (2025).

67. Chaurasia, A., Culurciello, E.: LinkNet: Exploiting encoder representations for efficient semantic segmentation. In: 2017 IEEE Visual Communications and Image Processing (VCIP). pp. 1–4. IEEE (2017).

68. Deng, J., Dong, W., Socher, R., Li, L.-J., Li, K., Fei-Fei, L.: ImageNet: A large-scale hierarchical image database. In: 2009 IEEE Conference on Computer Vision and Pattern Recognition. pp. 248–255. IEEE (2009).

69. Rokuss, M., Kirchhoff, Y., Roy, S., Kovacs, B., Ulrich, C., Wald, T., Zenk, M., Denner, S., Isensee, F., Vollmuth, P., Kleesiek, J., Maier-Hein, K.: Longitudinal segmentation of MS lesions via temporal Difference Weighting, http://arxiv.org/abs/2409.13416, (2024).

70. Ren, J., Hochreuter, K., Rasmussen, M.E., Kallehauge, J.F., Korreman, S.: Gradient Map-Assisted Head and Neck Tumor Segmentation: A Pre-RT to Mid-RT Approach in MRI-Guided Radiotherapy. In: Wahid, K.A., Dede, C., Naser, M.A., and Fuller, C.D. (eds.) LNCS. Springer, Cham (2025).

71. Ni, J., Yao, Q., Liu, Y., Qi, H.: A Coarse-to-Fine Framework for Mid-Radiotherapy Head and Neck Cancer MRI Segmentation. In: Wahid, K.A., Dede, C., Naser, M.A., and Fuller, C.D. (eds.) LNCS. Springer, Cham (2025).

72. Gao, Z., Tan, C., Wu, L., Li, S.Z.: SimVP: Simpler yet Better Video Prediction, http://arxiv.org/abs/2206.05099, (2022).

73. Armato, S.G., 3rd, Drukker, K., Hadjiiski, L.: AI in medical imaging grand challenges: translation from competition to research benefit and patient care. Br. J. Radiol. 96, 20221152 (2023).

74. Wang, J.-S., Wang, H.-J., Qian, H.-L.: Biological effects of radiation on cancer cells. Mil. Med. Res. 5, 20 (2018).

75. Joint Head and Neck Radiation Therapy-MRI Development Cooperative, MR-Linac Consortium Head and Neck Tumor Site Group: Longitudinal diffusion and volumetric kinetics of head and neck cancer magnetic resonance on a 1.5 T MR-linear accelerator hybrid system: A prospective R-IDEAL stage 2a imaging biomarker characterization/pre-qualification study. Clin. Transl. Radiat. Oncol. 42, 100666 (2023).

76. Ding, Y., Hazle, J.D., Mohamed, A.S.R., Frank, S.J., Hobbs, B.P., Colen, R.R., Gunn, G.B., Wang, J., Kalpathy-Cramer, J., Garden, A.S., Lai, S.Y., Rosenthal, D.I., Fuller, C.D.: Intravoxel incoherent motion imaging kinetics during chemoradiotherapy for human papillomavirus-associated squamous cell carcinoma of the oropharynx: preliminary results from a prospective pilot study: IVIM Kinetics during Chemoradiotherapy. NMR Biomed. 28, 1645–1654 (2015).

77. Wahid, K.A., Cardenas, C.E., Marquez, B., Netherton, T.J., Kann, B.H., Court, L.E., He, R., Naser, M.A., Moreno, A.C., Fuller, C.D., Fuentes, D.: Evolving Horizons in Radiation Therapy Auto-Contouring: Distilling Insights, Embracing Data-Centric Frameworks, and Moving Beyond Geometric Quantification. Adv Radiat Oncol. 9, 101521 (2024).

78. Fang, Y., Wang, J., Ou, X., Ying, H., Hu, C., Zhang, Z., Hu, W.: The impact of training sample size on deep learning-based organ auto-segmentation for head-and-neck patients. Phys. Med. Biol. 66, (2021). https://doi.org/10.1088/1361-6560/ac2206.

79. Rodríguez Outeiral, R., Bos, P., van der Hulst, H.J., Al-Mamgani, A., Jasperse, B., Simões, R., van der Heide, U.A.: Strategies for tackling the class imbalance problem of oropharyngeal primary tumor segmentation on magnetic resonance imaging. Phys Imaging Radiat Oncol. 23, 144–149 (2022).

80. Cardenas, C.E., Blinde, S.E., Mohamed, A.S.R., Ng, S.P., Raaijmakers, C., Philippens, M., Kotte, A., Al-Mamgani, A.A., Karam, I., Thomson, D.J., Robbins, J., Newbold, K., Fuller, C.D., Terhaard, C.: Comprehensive Quantitative Evaluation of Variability in Magnetic Resonance-Guided Delineation of Oropharyngeal Gross Tumor Volumes and High-Risk Clinical Target Volumes: An R-IDEAL Stage 0 Prospective Study. Int. J. Radiat. Oncol.

Biol. Phys. 113, 426–436 (2022).

81. Grégoire, V., Evans, M., Le, Q.-T., Bourhis, J., Budach, V., Chen, A., Eisbruch, A., Feng, M., Giralt, J., Gupta, T., Hamoir, M., Helito, J.K., Hu, C., Hunter, K., Johansen, J., Kaanders, J., Laskar, S.G., Lee, A., Maingon, P., Mäkitie, A., Micciche', F., Nicolai, P., O'Sullivan, B., Poitevin, A., Porceddu, S., Składowski, K., Tribius, S., Waldron, J., Wee, J., Yao, M., Yom, S.S., Zimmermann, F., Grau, C.: Delineation of the primary tumour Clinical Target Volumes (CTV-P) in laryngeal, hypopharyngeal, oropharyngeal and oral cavity squamous cell carcinoma: AIRO, CACA, DAHANCA, EORTC, GEORCC, GORTEC, HKNPCSG, HNCIG, IAG-KHT, LPRHHT, NCIC CTG, NCRI, NRG Oncology, PHNS, SBRT, SOMERA, SRO, SSHNO, TROG consensus guidelines. Radiother. Oncol. 126, 3–24 (2018).

82. Grégoire, V., Ang, K., Budach, W., Grau, C., Hamoir, M., Langendijk, J.A., Lee, A., Le, Q.-T., Maingon, P., Nutting, C., O'Sullivan, B., Porceddu, S.V., Lengele, B.: Delineation of the neck node levels for head and neck tumors: a 2013 update. DAHANCA, EORTC, HKNPCSG, NCIC CTG, NCRI, RTOG, TROG consensus guidelines. Radiother. Oncol. 110, 172–181 (2014).

83. Ren, J., Eriksen, J.G., Nijkamp, J., Korreman, S.S.: Comparing different CT, PET and MRI multi-modality image combinations for deep learning-based head and neck tumor segmentation. Acta Oncol. 60, 1399–1406 (2021).

84. Zhao, Y., Wang, X., Phan, J., Chen, X., Lee, A., Yu, C., Huang, K., Court, L.E., Pan, T., Wang, H., Wahid, K.A., Mohamed, A.S.R., Naser, M., Fuller, C.D., Yang, J.: Multi-modal segmentation with missing image data for automatic delineation of gross tumor volumes in head and neck cancers. Med. Phys. 51, 7295–7307 (2024).

85. Wahid, K.A., Ahmed, S., He, R., van Dijk, L.V., Teuwen, J., McDonald, B.A., Salama, V., Mohamed, A.S.R., Salzillo, T., Dede, C., Taku, N., Lai, S.Y., Fuller, C.D., Naser, M.A.: Evaluation of deep learning-based multiparametric MRI oropharyngeal primary tumor auto-segmentation and investigation of input channel effects: Results from a prospective imaging registry. Clin Transl Radiat Oncol. 32, 6–14 (2022).

86. Moawad, A.W., Janas, A., Baid, U., Ramakrishnan, D., Saluja, R., Ashraf, N., Jekel, L., Amiruddin, R., Adewole, M., Albrecht, J., Anazodo, U., Aneja, S., Anwar, S.M., Bergquist, T., Calabrese, E., Chiang, V., Chung, V., Conte, G.M.M., Dako, F., Eddy, J., Ezhov, I., Familiar, A., Farahani, K., Iglesias, J.E., Jiang, Z., Johanson, E., Kazerooni, A.F., Kofler, F., Krantchev, K., LaBella, D., Van Leemput, K., Li, H.B., Linguraru, M.G., Link, K.E., Liu, X., Maleki, N., Meier, Z., Menze, B.H., Moy, H., Osenberg, K., Piraud, M., Reitman, Z., Shinohara, R.T., Tahon, N.H., Nada, A., Velichko, Y.S., Wang, C., Wiestler, B., Wiggins, W., Shafique, U., Willms, K., Avesta, A., Bousabarah, K., Chakrabarty, S., Gennaro, N., Holler, W., Kaur, M., LaMontagne, P., Lin, M., Lost, J., Marcus, D.S., Maresca, R., Merkaj, S., Nada, A., Pedersen, G.C., von Reppert, M., Sotiras, A., Teytelboym, O., Tillmans, N., Westerhoff, M., Youssef, A., Godfrey, D., Floyd, S., Rauschecker, A., Villanueva-Meyer, J., Pflüger, I., Cho, J., Bendszus, M., Brugnara, G., Cramer, J., Perez-Carillo, G.J.G., Johnson, D.R., Kam, A., Kwan, B.Y.M., Lai, L., Lall, N.U., Memon, F., Patro, S.N., Petrovic, B., So, T.Y., Thompson, G., Wu, L., Schrickel, E.B., Bansal, A., Barkhof, F., Besada, C., Chu, S., Druzgal, J., Dusoi, A., Farage, L., Feltrin, F., Fong, A., Fung, S.H., Gray, R.I., Ikuta, I., Iv, M., Postma, A.A., Mahajan, A., Joyner, D., Krumpelman, C., Letourneau-Guillon, L., Lincoln, C.M., Maros, M.E., Miller, E., Morón, F., Nimchinsky, E.A., Ozsarlak, O., Patel, U., Rohatgi, S., Saha, A., Sayah, A., Schwartz, E.D., Shih, R., Shiroishi, M.S., Small, J.E., Tanwar, M., Valerie, J., Weinberg, B.D., White, M.L., Young, R., Zohrabian, V.M., Azizova, A., Brüßeler, M.M.T., Fehringer, P., Ghonim, M., Ghonim, M., Gkampenis, A., Okar, A., Pasquini, L., Sharifi, Y., Singh, G., Sollmann, N., Soumala, T., Taherzadeh, M., Yordanov, N., Vollmuth, P., Foltyn-Dumitru, M., Malhotra, A., Abayazeed, A.H., Dellepiane, F., Lohmann, P., Pérez-García, V.M., Elhalawani, H., Al-Rubaiey, S., Armindo, R.D., Ashraf,

K., Asla, M.M., Badawy, M., Bisschop, J., Lomer, N.B., Bukatz, J., Chen, J., Cimflova, P., Corr, F., Crawley, A., Deptula, L., Elakhdar, T., Shawali, I.H., Faghani, S., Frick, A., Gulati, V., Haider, M.A., Hierro, F., Dahl, R.H., Jacobs, S.M., Hsieh, K.-C.J., Kandemirli, S.G., Kersting, K., Kida, I., Kollia, S., Koukoulithras, I., Li, X., Abouelatta, A., Mansour, A., Maria-Zamfirescu, R.-C., Marsiglia, M., Mateo-Camacho, Y.S., McArthur, M., McDonnell, O., McHugh, M., Moassefi, M., Morsi, S.M., Muntenu, A., Nandolia, K.K., Naqvi, S.R., Nikanpour, Y., Alnoury, M., Nouh, A.M.A., Pappafava, F., Patel, M.D., Petrucci, S., Rawie, E., Raymond, S., Roohani, B., Sabouhi, S., Sanchez-Garcia, L.M., Shaked, Z., Suthar, P.P., Altes, T., Isufi, E., Dhermesh, Y., Gass, J., Thacker, J., Tarabishy, A.R., Turner, B., Vacca, S., Vilanilam, G.K., Warren, D., Weiss, D., Willms, K., Worede, F., Yousry, S., Lerebo, W., Aristizabal, A., Karargyris, A., Kassem, H., Pati, S., Sheller, M., Bakas, S., Rudie, J.D., Aboian, M.: The Brain Tumor Segmentation - Metastases (BraTS-METS) Challenge 2023: Brain Metastasis Segmentation on Pre-treatment MRI. ArXiv. (2024).

87. Raina, V., Molchanova, N., Graziani, M., Malinin, A., Muller, H., Cuadra, M.B., Gales, M.: Tackling Bias in the Dice Similarity Coefficient: Introducing nDSC for White Matter Lesion Segmentation, http://arxiv.org/abs/2302.05432, (2023).

88. Sherer, M.V., Lin, D., Elguindi, S., Duke, S., Tan, L.-T., Cacicedo, J., Dahele, M., Gillespie, E.F.: Metrics to evaluate the performance of auto-segmentation for radiation treatment planning: A critical review. Radiother. Oncol. 160, 185–191 (2021).

89. McCullum, L., Wahid, K.A., Marquez, B., Fuller, C.D.: OAR-Weighted Dice Score: A spatially aware, radiosensitivity aware metric for target structure contour quality assessment. Use Comput Radiat Ther. 2024, 755–758 (2024).

90. Wahid, K.A., Sahlsten, J., Jaskari, J., Dohopolski, M.J., Kaski, K., He, R., Glerean, E., Kann, B.H., Mäkitie, A., Fuller, C.D., Naser, M.A., Fuentes, D.: Harnessing uncertainty in radiotherapy auto-segmentation quality assurance. Phys Imaging Radiat Oncol. 29, 100526 (2024).

91. Wahid, K.A., Kaffey, Z.Y., Farris, D.P., Humbert-Vidan, L., Moreno, A.C., Rasmussen, M., Ren, J., Naser, M.A., Netherton, T.J., Korreman, S., Balakrishnan, G., Fuller, C.D., Fuentes, D., Dohopolski, M.J.: Artificial intelligence uncertainty quantification in radiotherapy applications − A scoping review. Radiotherapy and Oncology. 201, (2024). https://doi.org/10.1016/j.radonc.2024.110542.

92. El-Habashy, D.M., Wahid, K.A., He, R., McDonald, B., Mulder, S.J., Ding, Y., Salzillo, T., Lai, S.Y., Christodouleas, J., Dresner, A., Wang, J., Naser, M.A., Fuller, C.D., Mohamed, A.S.R., Joint Head and Neck Radiation Therapy-MRI Development Cooperative: Dataset of weekly intra-treatment diffusion weighted imaging in head and neck cancer patients treated with MR-Linac. Sci. Data. 11, 487 (2024).

93. Andrearczyk, V., Oreiller, V., Jreige, M., Vallières, M., Castelli, J., Elhalawani, H., Boughdad, S., Prior, J.O., Depeursinge, A.: Overview of the HECKTOR Challenge at MICCAI 2020: Automatic Head and Neck Tumor Segmentation in PET/CT. In: Head and Neck Tumor Segmentation. pp. 1–21. Springer International Publishing (2021).

94. Andrearczyk, V., Oreiller, V., Boughdad, S., Rest, C.C.L., Elhalawani, H., Jreige, M., Prior, J.O., Vallières, M., Visvikis, D., Hatt, M., Depeursinge, A.: Overview of the HECKTOR Challenge at MICCAI 2021: Automatic Head and Neck Tumor Segmentation and Outcome Prediction in PET/CT Images. In: Head and Neck Tumor Segmentation and Outcome Prediction. pp. 1–37. Springer International Publishing (2022).